\documentclass[12pt]{article}
\pdfoutput=1
\usepackage{bm}
\usepackage{amsmath}
\numberwithin{equation}{section}
\usepackage{amssymb}
\usepackage{graphicx}
\usepackage{float}
\usepackage{bbold}
\usepackage{multirow}
\usepackage{enumerate}
\usepackage{dynkin-diagrams}
\usepackage{ulem}
\usepackage{color}

\usepackage[utf8]{inputenc}
\usepackage[hidelinks]{hyperref}
\tikzset{/Dynkin diagram, root radius=0.1cm, edge length=0.9cm}

\begin{document}
\begin{titlepage}
		\begin{flushright}
			TIT/HEP-690 \\
			June,  2022
		\end{flushright}
		\vspace{0.5cm}
		\begin{center}
			{\Large \bf
				ODE/IM correspondence and supersymmetric affine Toda field equations
			}
			\lineskip.75em
			\vskip 2.5cm
			{\large  Katsushi Ito and Mingshuo Zhu }
			\vskip 2.5em
			{\normalsize\it Department of Physics,\\
				Tokyo Institute of Technology\\
				Tokyo, 152-8551, Japan}
			\vskip 3.0em
		\end{center}
		\begin{abstract}
We study the linear differential system associated with the supersymmetric affine Toda field equations for affine Lie superalgebras, which has a purely odd simple root system. For an affine Lie algebra, the linear problem modified by conformal transformation leads to an ordinary differential equation (ODE) that provides the functional relations in the integrable models. 
This is known as the ODE/IM correspondence. 
For the affine Lie superalgebras, the linear equations modified by a superconformal transformation are shown to reduce to a couple of ODEs for each bosonic subalgebra. In particular, for $osp(2,2)^{(2)}$, the corresponding ODE becomes the second-order ODE with squared potential, which is related to the ${\cal N}=1$ supersymmetric minimal model via the ODE/IM correspondence. We also find ODEs for classical affine Lie superalgebras with purely odd simple root systems.
		\end{abstract}
\end{titlepage}
	\baselineskip=0.7cm
	
	%\maketitle
	\numberwithin{equation}{section}
	\numberwithin{figure}{section}
	\numberwithin{table}{section}
	%\tableofcontents
\section{Introduction}
The relation between classical and quantum integrable models has recently attracted much attention. The classical integrable systems are realized as a Lax pair of the linear differential equations. For a particular class of linear problems, the global structure of the solutions determines the spectral of the quantum integrable models. A typical example is a relation between connection coefficients of the modified sinh-Gordon equation and the Q-functions in the quantum sine(h)-Gordon model \cite{Lukyanov:2010rn}.
This correspondence has been generalized to modified affine Toda field equations \cite{Dorey:2012bx,Ito:2013aea,Adamopoulou:2014fca,Ito:2018wgj},
which plays also an important role in studying the gluon-scattering amplitudes at strong coupling \cite{Alday:2010vh,Hatsuda:2010cc}.

The conformal limit of the above linear differential system reduces to the single ordinary linear differential equation of a holomorphic variable. The ODE becomes the Schr\"odinger equation for the sine(h)-Gordon model, which has appeared in the study of the ODE/IM correspondence between the second-order ODE and the six-vertex model or related conformal field theory (CFT) \cite{Dorey:1998pt,Bazhanov:1998wj}.

The main purpose of this paper is to explore the correspondence between the ODEs and CFTs (or IMs) to make a complete dictionary. In particular, CFT with supersymmetry is a subject of the present work. 
It turns out that precise correspondence depends on the potential term of the ODE. For the Schr\"odinger case, where the ODE takes the form $(-\frac{d^2}{dx^2}+P(x))\psi(x)=0$, the generic polynomial potential $P(x)$ in the minimal chamber leads to the homogeneous sine-Gordon model \cite{Ito:2018eon} (See also \cite{Ito:2021boh,Ito:2021sjo} for the higher-order ODE). However, for monomial potential $P(x)=x^{2M}-E$, it corresponds to the Virasoro minimal model. Moreover, when the potential term becomes the $K$-th power of the monomial potential $P(x)=(x^{2M/K}-E)^K$, the theory becomes the $SU(2)$ coset model $SU(2)_K \times SU(2)_L/SU(2)_{K+L}$, which has been also generalized to the higher-order ODE \cite{Dorey:2006an} (see also \cite{Lukyanov:2006gv}). In particular, for $K=2$, the chiral algebra of the CFT enhances to ${\cal N}=1$ superconformal symmetry. 
A detailed analysis for $K=2$ has been done in \cite{Babenko:2017fmu}, where they observed the correspondence between the Suzuki equation \cite{Suzuki:1998ve,Dunning:2002tt} and the WKB periods. In \cite{Kulish:2004ap,Kulish:2005qc}, the quantization  of ${\cal N}=1$ supersymmetric scalar Lax operator based on an affine Lie superalgebra  $osp(2,2)^{(2)}$ leads to the supersymmetric integrable structure of ${\cal N}=1$ minimal model, which provides a supersymmetric generalization of \cite{Bazhanov:1994ft}. The ODE/IM correspondence with supersymmetry would also provide a method to study the integrability  of superstrings on $AdS_{5}\times S^{5}$ associated with supercoset model $PSU(2,2|4)/SO(4,1)\times SO(5)$ \cite{Bena:2003wd}.

The conformal limit of the linear problems associated with affine Toda field equations and the ODE/IM correspondence have been studied in \cite{Sun:2012xw,Masoero:2015lga,Ito:2020htm}.	
In this paper,  we explore the manifestly supersymmetric formulation of the ODE/IM correspondence based on the affine Lie superalgebras. In particular, we focus on affine Toda field theory based on the algebra with a purely fermionic simple root system, which admits ${\cal N}=1$ superspace formulation. Such affine Lie superalgebras are classified as 
$A(n|n)^{(1)}, D(n|n-1)^{(1)}, B(n|n)^{(1)}, C(2)^{(2)}, A(n|n)^{(2)},D(n|n)^{(2)}, D(2|1;\alpha)^{(2)}$
\cite{serganova,Leites:1985hh}. The supersymmetric linear problems are formulated for these algebras. An affine Lie superalgebra contains two bosonic subalgebras. When we set the fermionic part of the superfields to zero, one obtains the linear problem for the bosonic subalgebras. Taking the conformal limit, one further obtains the ODE for the affine Lie superalgebra. For the untwisted case, we obtain two decoupled ODEs for each subalgebra with $K=1$. However, for the twisted case, we find a nontrivial linear system. Especially for $C(2)^{(2)}=osp(2|2)^{(2)}$, it reproduces the ODE with $K=2$. We also explore the ODEs for other affine Lie superalgebras with higher ranks except for $D(2|1;\alpha)^{(2)}$. A detailed analysis of the ODE/IM correspondence will be studied in a separate paper.

This paper is organized as follows. In Section \ref{Section 2}, we generalize the affine Toda field equations to $\mathcal{N}=1$ supersymmetric case. We define the linear problem associated with the affine Toda field equations and derive the conformal limit of the problem. We also discuss the reduction of the super Lax operator to the bosonic part, which defines the bosonic Lax operator.

The detailed analysis to $C(2)^{(2)}$ algebra is carried out in Section \ref{Section 3}, where we show the bosonic reduction of the  Lax operator leads to an ODE with the squared potential term. In Section \ref{Section 4}, we summarize the ODE structure for other affine Lie superalgebras with purely fermionic simple root systems.
Section 5 is devoted to conclusions and discussion.

\section{Linear problem and supersymmetric affine Toda field equations}\label{Section 2}
In this section, we study the linear differential equations associated with the affine Toda field equation based on an affine Lie superalgebra. Let ${\mathfrak g}$ be a Lie superalgebra, which is generated by $H^i$ ($i=1,\dots, r$) and $E_{\alpha}$ ($\alpha \in \Delta$) \cite{Kac}. Here $H^i$ are the generators of the Cartan subalgebra, and $r$ is the rank of ${\mathfrak g}$. The set of roots $\Delta$ of ${\mathfrak g}$ is decomposed into the sets of even and odd roots, whose corresponding generators are commuting (bosonic) and anti-commuting (fermionic). The roots are expressed in terms of the simple roots denoted by $\{\alpha_1,\dots, \alpha_r\}$. Let ${\cal F}$ be a set of simple roots whose generators are fermionic and ${\cal B}$ a set of simple roots corresponding to the bosonic generators. In contrast to Lie algebra, a Lie superalgebra admits a different simple root system. 
 
For a Lie superalgebra ${\mathfrak g}$, one can define an affine Lie superalgebra $\hat{\mathfrak g}$ \cite{FSS}. The untwisted affine Lie algebra is defined as ${\mathfrak g}^{(1)}=	{\mathbb C}[t,t^{-1}]\otimes {\mathfrak g}\oplus {\mathbb C}k$, where ${\mathbb C}[t,t^{-1}]$ denotes the Laurent series in $t$ and $k$ represents the level of the algebra. 
The simple root system is extended by adding a root $\alpha_0=-\theta$, where $\theta$ denotes the highest root. The twisted affine Lie algebra ${\mathfrak g}^{(m)}$ ($m\geq 2$) is defined by an outer automorphism $\tau$ of ${\mathfrak g}$ of degree $m$: $\tau^m=1$ \cite{vanderleur}. ${\mathfrak g}$ is decomposed into the subspace ${\mathfrak g}_k$ ($k=0,\dots,m-1$) with $\tau$ eigenvalue $e^{2\pi i k\over m}$. Then ${\mathfrak g}^{(m)}$ is defined by ${\mathfrak g}^{(m)}=(\oplus_{n\in {\mathbb Z}}\oplus_{k=0}^{m-1} {\mathfrak g}_k\otimes t^{n+k/m})\oplus {\mathbb C}k$. In this paper, we focus on $m=2$ case, where the Dynkin diagram is obtained from the one of ${\mathfrak g}_0$ by adding the lowest weight in ${\mathfrak g}_1$ as an extended root \cite{FSS}.

Now we define ${\cal N}=1$ supersymmetric affine Toda field equation based on an affine Lie superalgebra $\hat{\mathfrak g}$ \cite{Olshanetsky:1982sb}. To realize ${\cal N}=1$ supersymmetry, we introduce ${\cal N}=1$ superspace with complex coordinates $(z,\theta,\bar{z},\bar{\theta})$, where $\theta$ and  $\bar{\theta}$ are Grassmann variables. The covariant superderivative is defined by
	\begin{equation}
		D=\frac{\partial}{\partial\theta}+\theta\partial,\quad \bar{D}=\frac{\partial}{\partial\bar{\theta}}+\bar{\theta}\bar{\partial},
	\end{equation}
	\begin{equation}
		D^{2}=\partial,\quad \bar{D}^{2}=\bar{\partial},\quad \lbrace D,\bar{D}\rbrace=0.
	\end{equation}
We also introduce an $r$-component scalar superfield by
\begin{equation}\label{superfield}
		\Phi(z,\bar{z},\theta,\bar{\theta})=\phi(z,\bar{z})+i\theta\eta(z,\bar{z})+i\bar{\theta}\bar{\eta}(z,\bar{z})+\theta\bar{\theta}F(z,\bar{z}),
\end{equation}
where $\phi(z,\bar{z})$, $F(z,\bar{z})$ are bosonic fields, $\eta(z,\bar{z})$ and $\bar{\eta}(z,\bar{z})$ are fermionic fields. 

The action of Toda field theory for a Lie superalgebra ${\mathfrak g}$ is defined by
	\begin{equation}
		S[\Phi]=\int d^{2}zd^{2}\theta\bigg[\frac{1}{2} D\Phi\cdot \bar{D}\Phi-\frac{m^2}{\beta^{2}}\sum_{\alpha_{i}\in \mathcal{F}}\exp(\beta\alpha_{i}\cdot\Phi)-\frac{m^2}{\beta^{2}}\theta\bar{\theta}\sum_{\alpha_{i}\in \mathcal{B}}\exp(\beta\alpha_{i}\cdot\Phi)\bigg],
		\label{eq:action_supertoda}
	\end{equation}
where $\beta$ is a coupling constant, $m$ is  a mass parameter,   and $\alpha_{i}$ are the simple roots of ${\mathfrak g}$ \cite{Evans:1990qq,Komata:1990cb,Delduc:1991sg}. 
The symbol $\cdot$ denotes the inner product in the root space. The action  of the affine Toda field theory for an affine Lie superalgebra $\hat{\mathfrak g}$ is obtained by adding the potential associated with the root $\alpha_0$ to the Lagrangian, where the additional potential term is $-\frac{m^2}{\beta^2} \exp(\beta \alpha_0\cdot\Phi)$ for $\alpha_0$ odd and $-\frac{m^2}{\beta^2} \theta\bar{\theta}\exp(\beta \alpha_0\cdot\Phi)$  for $\alpha_0$ even. 

It can be seen that the action (\ref{eq:action_supertoda}) is not invariant under ${\cal N}=1$ supersymmetry due to the existence of simple bosonic roots. This paper will focus on the class of affine Toda field theories, which admits manifestly ${\cal N}=1$ supersymmetric formulation. The possible affine Lie superalgebras are classified in \cite{serganova}, which are  summarized in the Table \ref{tab:table_liesuperalgebras}. Here we omit the exceptional affine Lie superalgebra $D(2|1;\alpha)^{(1)}$ in the table because its matrix representation is rather complicated. See \cite{Andreev:1987xj}.
\begin{table}[tbh]
\begin{center}
\begin{tabular}{|l|c|c|}
\hline
affine Lie & Dynkin diagram & simple roots\\
 superalgebra & & $\alpha_1,\dots, \alpha_r,\alpha_{r+1}$ \\
\hline
$A(n|n)^{(1)}$ &  \dynkin[affine mark=t]A[1]{t.t.t} &$e_1-\delta_1,\delta_1-e_2, \dots, e_{n+1}-\delta_{n+1}$\\
& & $\delta_{n+1}-e_1$\\
\hline
$D(n|n-1)^{(1)}$ & & 
$e_1-\delta_1,\delta_1-e_2,\dots, \delta_{n-1}-e_n,\delta_{n-1}+e_n$\\
&\dynkin[fold,double edges,affine mark=t] D[1]{tt.ttt} & $-e_1-\delta_1$ \\
%\\
\hline
$D(2|1)^{(1)}$ & \dynkin [ply=4,double edges,affine mark=t] A[1]{ttt} & $e_1-\delta_1,\delta_1-e_2,\delta_1+e_2, -\delta_1-e_1$\\
\hline
$B(n|n)^{(1)}$& 
\dynkin[fold,double edges, affine mark=t] B[1]{tt.t.t*}
& 
$e_1-\delta_1,\delta_1-e_2,\dots, e_n-\delta_n, \delta_n$ \\
& & $-e_1-\delta_1$ \\
%\\
\hline
$C(2)^{(2)}$ & \dynkin [arrows=false, affine mark=*] A[1]{*}
%\dynkin A{**}
& $\delta_1,-\delta_1$
\\
\hline
$A(2n|2n)^{(2)}$  &  \dynkin[affine mark=*] D[2]{tt.t*}&
$\delta_1,e_{n+1}-\delta_1,\delta_2-e_{n+1},\dots, e_1-\delta_{n+1},-e_1$
\\
\hline
$A(1|1)^{(2)}$ & \dynkin[affine mark=t] D[2]{tt}& $e_1+\delta_1,\delta_1-e_1,-\delta_1-e_1$\\
\hline
{\small $A(2n+1| 2n+1)^{(2)}$}  & \dynkin[fold,double edges,affine mark=t]D[1]{tt.t.ttt} & $e_n+\delta_1,-e_n+\delta_1,\-\delta_1+e_{n-1},\dots, -\delta_{n-1}+e_1,$\\
($n\geq 1$) & & $ \delta_n-e_1,-\delta_n-e_1$\\
\hline
$D(n|n)^{(2)}$ &  \dynkin[affine mark=*] D[2]{tt.t*} & $\delta_1-e_1,e_1-\delta_2,\dots, e_{n-1}-\delta_n, \delta_n,-\delta_1$\\
\hline
\end{tabular}
\end{center}
\caption{Dynkin diagrams of classical affine Lie algebras with purely fermionic simple root system, where a black dot represents the odd simple root with non-zero length and a cross dot means the odd simple root with zero length.\cite{FSS}. $e_i$ and $\delta_i$ are orthonormal basis with positive (negative) norm. For $A(2n|2n)^{(2)}$ and $A(2n+1|2n+1)^{(2)}$, $\alpha_0:=\alpha_{n+1}$. For other algebras, $\alpha_0:=\alpha_{r+1}$.}
\label{tab:table_liesuperalgebras}
\end{table}

Now for affine Lie superalgebras with purely odd simple root systems, the affine Toda field equations are obtained as the equations of motion from the Lagrangian and take the form of
	\begin{equation}%\label{ssge}
		D\bar{D}\Phi+\frac{m^2}{\beta}\sum_{i=1}^{r}\alpha_{i}\exp(\beta \alpha_{i}\cdot\Phi)+\frac{m^2}{\beta}\alpha_{0}\exp(\beta \alpha_{0}\cdot\Phi)=0.
		\label{eq:affinetodafieldeq1}
	\end{equation}

\subsection{Modified supersymmetric affine Toda field equations}
The Toda field equation (\ref{eq:affinetodafieldeq1}) (without the third term) is invariant under the superconformal transformation:
\begin{align}
		\begin{split}
			&z'(z,\theta)=f(z)+\theta\epsilon(z)g(z),\quad \theta'=\epsilon(z)+\theta+\theta g(z),\\
			&\bar{z}'(\bar{z},\bar{\theta})=\bar{f}(\bar{z})+\bar{\theta}\bar{\epsilon}(\bar{z})\bar{g}(\bar{z}),\quad \bar{\theta}'=\bar{\epsilon}(\bar{z})+\bar{\theta}+\bar{\theta} \bar{g}(\bar{z}),\\
			&\Phi'(z,\theta,\bar{z},\bar{\theta})=\Phi(z',\theta',\bar{z}',\bar{\theta}')+\frac{1}{\beta}\mu\log(D\theta'\bar{D}\bar{\theta}'),\\
		\end{split}
		\label{eq:superconformal}
\end{align}
where $\mu=\frac{1}{2}\sum_{i=1}^{r}\mu_{i}$ is half the sum of the fundamental weights $\mu_i$ satisfying $\mu_i\cdot \alpha_j=\delta_{ij}$ ($i,j=1,\dots, r$). Under the superconformal transformation (\ref{eq:superconformal}),  
the super covariant derivative transforms as
	\begin{equation}
		D=(D\theta')D'+(Dz'-\theta' D\theta')\partial'.
	\end{equation}
The conformal transformation is defined by $Dz'=\theta D\theta'$, which determines $g(z)$ as $(\partial f)^{1/2}$.
Under the conformal transformation, the affine Toda field equation \eqref{eq:affinetodafieldeq1} is modified as
\begin{equation}
		D\bar{D}\Phi+\frac{m^2}{\beta}\sum_{i=1}^{r}\alpha_{i}\exp(\beta \alpha_{i}\cdot\Phi)+\frac{m^2}{\beta}p(z)\bar{p}(\bar{z})\alpha_{0}\exp(\beta \alpha_{0}\cdot\Phi)=0,
		\label{eq:modifiedaftfeq}
\end{equation}
	where 
\begin{align}
	p(z)=(\partial f)^{\frac{h}{2}}, \quad \bar{p}(\bar{z})=(\bar{\partial}\bar{f})^{\frac{h}{2}},
\end{align}
and $h$ is defined by
\begin{align}
\mu\cdot \alpha_0&=-{h-1\over2}.
\end{align}
When the extended root $\alpha_0$ is expanded in terms of simple roots as $\alpha_0=-\sum_{i=1}^{r}n_i \alpha_i$, one finds
$h=\sum_{i=1}^r n_i+1$, i.e. the Coxeter number of ${\frak g}$.
The values of $h$ for $\hat{\frak g}$ are listed in Table \ref{tab:coxeter1}.
\begin{table}[tbh]
\begin{tabular}{|c|c|c|c|c|c|c|}
\hline
$A(n|n)^{(1)}$ & $D(n|n-1)^{(1)}$ & $B(n|n)^{(1)}$ & $C(2)^{(2)}$ 
& $A(2n|2n)^{(2)}$ & {\small $A(2n-1|2n-1)^{(2)}$} & $D(n|n)^{(2)}$ \\
\hline
$2n+2$ &$4n-4$ & $4n$ & $2$ & $2n+1$ &$4n-2$ & $2n$ \\
\hline
\end{tabular}
\caption{List of the Coxeter number $h$ of affine Lie superalgebras in table \ref{tab:table_liesuperalgebras}. }
\label{tab:coxeter1}
\end{table}
The modified affine Toda field equation (\ref{eq:modifiedaftfeq}) is not manifestly supersymmetric due to $p(z)$ and $\bar{p}(\bar{z})$ in the complex coordinates $(z,\bar{z})$. However, using the conformal transformation, it can be transformed in the form (\ref{eq:affinetodafieldeq1}) which is supersymmetric.
	
\subsection{Super Lax operator}
Now we express the modified affine Toda field equation (\ref{eq:modifiedaftfeq}) in the Lax form.
Let $E_{\alpha_{i}}$, $E_{-\alpha_{i}}$, $H$ ($i=0,1,\dots, r$) be the  basis of $\hat{\frak g}$ satisfying the (anti-)commutation relations:
\begin{align}\label{eq:superalgebra}
\begin{split}
    &\{ E_{\alpha_{i}}, E_{-\alpha_{i}}\}=\epsilon_{i}\alpha_{i}\cdot H,\\ [\alpha_{i}\cdot H, E_{\alpha_{j}}]=(\alpha_i\cdot \alpha_j)&E_{\alpha_{j}},\quad [\alpha_{i}\cdot H,E_{-\alpha_{j}}]=-(\alpha_i\cdot\alpha_j)E_{-\alpha_{j}},\\
    \end{split}
\end{align}
where $\epsilon_{i}$ is a normalization factor chosen to be $\pm 1$, which depends on the representation of $\hat{\mathfrak g}$. 
See table \ref{tab:normalization} for examples.

We define the Lax operators by
\begin{align}\label{eq:superLax}
		&\mathcal{L}_{F}=D-A_{\theta}=D-\beta D\Phi\cdot H-m e^{\lambda}(\sum_{i=1}^{r}E_{\alpha_{i}}+p(z)E_{\alpha_{0}}),\\
		&\bar{\mathcal{L}}_{F}=\bar{D}-A_{\bar{\theta}}=\bar{D}+m e^{-\lambda}(\sum_{i=1}^{r}e^{\beta\alpha_{i}\cdot\Phi}\epsilon_{i}E_{-\alpha_{i}}+\bar{p}(\bar{z})e^{\beta\alpha_{0}\cdot\Phi}\epsilon_{0}E_{-\alpha_{0}}). \nonumber
\end{align}
Here we introduced the spectral parameter $\lambda$.
The modified affine Toda field equation	\eqref{eq:modifiedaftfeq} can be rewritten as the flatness condition of the Lax operators:
	\begin{equation}
		F_{\theta\bar{\theta}}=DA_{\bar{\theta}}+\bar{D}A_{\theta}-\lbrace A_{\theta},A_{\bar{\theta}}\rbrace=0
	\end{equation}
The flatness condition is the compatibility of the two supersymmetric 
linear problems
\begin{align}
    {\cal L}_F \Psi(z,\bar{z},\theta,\bar{\theta})=0,\quad
    \bar{\cal L}_F\Psi(z,\bar{z},\theta,\bar{\theta})=0.
    \label{eq:linearprob}
\end{align}
We can write the Lax operators \eqref{eq:superLax} in a symmetric manner by using the gauge transformation:
\begin{align}
    A_\theta& \rightarrow A^U_\theta=U A_\theta U^{-1} -UD U^{-1}, \quad
    A^U_{\bar{\theta}}=A_{\bar{\theta}} \rightarrow U A_{\bar{\theta}}U^{-1}-U \bar{D} U^{-1}, \quad \chi \rightarrow U\chi
    \label{eq:gaugetr}
\end{align}
with $U=\exp(\frac{\beta}{2} \alpha_i\cdot \Phi)$.

Let us consider the first equation in (\ref{eq:linearprob}) in the component formalism. Expanding $\Psi(z,\theta)=\Psi_{0}(z)+i\theta\Psi_{1}(z)$ 
and $D\Phi=i\eta+\theta \partial_z \phi$ (we omit the anti-holomorphic part in super field), one obtains a set of equations % and plunge it into \eqref{super Lax equation}.
\if0
	\begin{equation}
		[\partial_{\theta}+\theta\partial_{z}-(\partial_{\theta}+\theta\partial_{z})(\phi_{j}+i\theta\eta_{j})h_{j}-e^{\lambda}(\sum_{i=1}e_{i}+p(z,\theta)e_{0})](\mathcal{O}_{0}(z)+i\theta\mathcal{O}_{1}(z))=0
	\end{equation}
	Then we can obtain
\fi
\begin{align}
\begin{split}\label{explicit superfield}
		& \partial_{z}\Psi_{0}-\beta\partial_{z}\eta\cdot H\Psi_{0}-i[me^{\lambda}(\sum_{i=1}^{r}E_{\alpha_{i}}+p(z)E_{\alpha_{0}})-i\beta\eta\cdot H]\Psi_{1}=0,\\
		& i\Psi_{1}-i\beta\eta\cdot H\Psi_{0}-me^{\lambda}(\sum_{i=1}^{r}E_{\alpha_{i}}+p(z)E_{\alpha_{0}})\Psi_{0}=0.
	\end{split}
	\end{align}
Eliminating the fermionic field $\Psi_{1}$, one can find the equation ${\cal L}_B\Psi_0=0$ with
	\begin{align}
	\begin{split}
		\mathcal{L}_{B}=&\partial_{z}-\beta\partial_{z}\phi\cdot H\\-&[me^{\lambda}(\sum_{i=1}^{r}E_{\alpha_{i}}+p(z)E_{\alpha_{0}})-i\beta\eta\cdot H][me^{\lambda}(\sum_{i=1}^{r}E_{\alpha_{i}}+p(z)E_{\alpha_{0}})+i\beta\eta\cdot H].\\
		\end{split}
	\end{align}
When further setting $\eta=0$, one obtains the Lax operator
	\begin{equation}\label{eq:bosonicLaxoperator}
		\mathcal{L}_{B}=\partial_{z}-\beta\partial_{z}\phi\cdot H-m^{2}e^{2\lambda}(\sum_{i=1}^{r}E_{\alpha_{i}}+p(z)E_{\alpha_{0}})^{2}.
	\end{equation}
Note that ${\cal L}_B$ takes values in the bosonic subalgebra of $\hat{\frak g}$ since the squared term in ${\cal L}_B$ can be written as half the anti-commutator of odd generators.
Then the bosonic reduction of the linear problem makes sense as in the linear problem for bosonic subalgebra.
By taking the conformal limit further, one obtains the linear problem of a single holomorphic variable. Then it is possible to apply the ODE/IM correspondence to the linear problem.

\subsection{Conformal limit of the linear problem}
We discuss the conformal (or light-cone) limit of the linear problem
\eqref{eq:linearprob}, where we take the scaling such that the anti-holomorphic part becomes negligible.
%	It is better to determine the asymptotic behavior here. Since in \eqref{mssg}, only conformal transformation is permitted, we can focus on the bosonic part. The asymptotic behavior here follows the assumption in \cite{Ito Locke}. 
To simplify the problem, we take $p(z)$ as the monomial in $z$ of the form
\begin{align}
    p(z)&=z^{hM}-s^{hM}, \quad \bar{p}(\bar{z})=\bar{z}^{hM}-s^{hM}.
\end{align}
The asymptotic behavior of the solution of the affine Toda field equation \eqref{eq:modifiedaftfeq} is given as follows:
for $|z|\rightarrow\infty$, the potential part becomes dominant.
Then leading part of $\Phi(z,\bar{z})$ is
\begin{equation}
	\Phi(z,\bar{z},\theta,\bar{\theta})=\frac{2M\mu}{\beta}\log(z\bar{z})+\cdots .
\end{equation}
As $|z|\rightarrow0$, we can impose certain boundary condition. 
Here we assume that $\Phi$ can be expanded as
	\begin{equation}\label{4.3.9}
		\Phi(z,\bar{z})=g\log(z\bar{z})+%\text{Regular boson}+\text{Fermion}
		\cdots,
	\end{equation}
Here sub-leading term might includes fermionic part in a general setup, which is not covered  in the present work.
Now we take the conformal limit of the linear problem.
We consider the gauge transformation \eqref{eq:gaugetr} with
$U=\exp(\frac{2}{h}\log p(z) \mu\cdot H)$. The connection becomes
\begin{align}
    A^U_\theta&=\beta D\Phi \cdot H-\frac{2}{h}D\log p(z) \mu\cdot H
    +m e^{\lambda} p(z)^{\frac{1}{h}}(\sum_{i=1}^{r}E_{\alpha_{i}}+E_{\alpha_{0}}).
\end{align}
We rescale $z$, $\bar{z}$, $s$, $\bar{s}$, $\theta$ and $\bar{\theta}$ as
\begin{align}
    x&=(m e^{\lambda})^{\frac{2}{2M+1}}z, \quad E=s^{hM} (m e^{\lambda})^{\frac{2hM}{2M+1}}, \quad \tilde{\theta}=\theta (m e^{\lambda})^{-\frac{1}{2M+1}}, \\
    \bar{x}&=(m e^{-\lambda})^{\frac{2}{2M+1}}\bar{z}, \quad \bar{E}=\bar{s}^{hM} (m e^{-\lambda})^{\frac{2hM}{2M+1}}, \quad \tilde{\bar{\theta}}=\theta (m e^{-\lambda})^{-\frac{1}{2M+1}}, \nonumber 
\end{align}
and take the light-cone limit $\bar{z}\rightarrow 0$ with $\lambda\rightarrow \infty$. Then we take the limit $z\rightarrow 0$ with $x$, $E$, $\tilde{\theta}$ kept finite.
In this limit, the differential equation for
the super Lax operator $U{\cal L}_F U^{-1}=D-A^U_\theta$ reduces to the equation ${\cal L}'\Psi=0$, where
\begin{align}
    {\cal L}'&={\cal D}-\tilde{\theta}\frac{l\cdot H}{x} +\frac{2}{h}\mathcal{D}\log p(x,E) \mu\cdot H-p(x,E)^{\frac{1}{h}} (\sum_{i=1}^{r}E_{\alpha_{i}}+E_{\alpha_{0}}),
    \label{eq:superlax2}
\end{align}
with ${\cal D}=\frac{\partial}{ \partial \tilde{\theta}}+\tilde{\theta}\partial_x$, $l=\beta g$ and $p(x,E)=x^{hM}-E$.
Applying inverse gauge transformation with the gauge parameter $\exp( -{2\over h}\log p(x,E) \mu\cdot H)$, one can obtain the Lax operator
 \begin{align}
     {\cal L}={\cal D}-\tilde{\theta}\frac{l\cdot H}{x} -\left( \sum_{i=1}^r E_{\alpha_i} +p(x,E) E_{\alpha_0} \right),
     \label{eq:lax3}
 \end{align}
 which leads to 
 the bosonic Lax operator of the form
	\begin{equation}\label{eq:bosoniclpcl1}
		\mathcal{L}_{B}=\partial_{x}-\sum_{j=1}^{r}\frac{l_{j}\alpha_{j}\cdot H}{x}-(\sum_{i=1}^{r}E_{\alpha_{i}}+p(x,E)E_{\alpha_{0}})^{2} .
	\end{equation}
Here $l=\sum_i l_i \alpha_i$.
The \eqref{eq:bosoniclpcl1} provides the differential operator of the ODE for the affine Lie superalgebra $\hat{\frak g}$.
In the following sections, we will study the linear problem associated with affine Toda field equations using the explicit matrix representation.

The asymptotic solution of the linear problem for ${\cal L}'\Psi=0$ in \eqref{eq:superlax2} for large $|x|$ is found by the eigenvectors of the matrix $\Lambda_+=\sum_{i=1}^{r}E_{\alpha_i}+E_{\alpha_0}$, since in this region the ODE becomes of the form ${\cal D}\Psi-p(x,E)^{\frac{1}{h}}\Lambda_+\Psi=0$. 
Let us consider the representation of ${\mathfrak g}$ with dimension $N$. Let ${\boldsymbol \mu}_a$ ($a=1,\dots, N$) be the eigenvector of $\Lambda_+$ with eigenvalue $\mu_a$. 
Then the basis of the asymptotic solutions of ${\cal L}'\Psi=0$ is found to be
\begin{align}
    \psi'_a(x,\tilde{\theta})&= \exp\left(\mu_a^2\int^x p(t,E)^{\frac{2}{h}} dt\right) (1-ip(x,E) \mu_a \tilde{\theta}) {\boldsymbol \mu}_a+\cdots .
\end{align}
The asymptotic solution around $x=0$ is determined by $({\cal D}-\tilde{\theta}\frac{l\cdot H}{x})\chi=0$. Its basis is given by
\begin{align}
    \chi'_a(x,\tilde{\theta})&= x^{l \cdot \nu_a} {\boldsymbol \nu}_a+\cdots.
\end{align}
Here ${\boldsymbol \nu}_a$ are the weight vector of the representation with  weight $\nu_a$.
For the linear problem ${\cal L}\Psi=0$ with ${\cal L}$ given in \eqref{eq:lax3},
the basis of the solutions are obtained by the gauge transformation: for large 
$|x|$, 
\begin{align}\label{eq:asymptoticinfity}
    \psi_a(x,\tilde{\theta})&=\exp( -{2\over h}\log p(x,E) \mu\cdot H)\psi'_a(x,\tilde{\theta}), \quad a=1,\dots, N,
\end{align}
and for small $|x|$,
\begin{align}\label{eq:asymptoticorigin}
    \chi_a(x,\tilde{\theta})&=\exp( -{2\over h}\log p(x,E) \mu\cdot H)\chi'_a(x,\tilde{\theta}), \quad a=1,\dots, N.
\end{align}

Suppose that $\psi(x,\tilde{\theta})$ is the subdominant solution along the positive real axis, where we need to choose the eigenvalue such that the real part of $\mu_a^2$ is the most negative. Then one can define the Q-function by
\begin{align}\label{eq:Qfunction}
    \psi(x,\tilde{\theta})&=\sum_{a} Q_a (E,l)\chi_a(x,\tilde{\theta}).
\end{align}
Employing the symmetry under the Symanzik rotation
$(x,\tilde{\theta},E)\rightarrow (\omega x, \omega^{\frac{1}{2}}\tilde{\theta}, \omega^{hM}E)$ with $\omega^{h(2M+1)}=1$, the Wronskians of the asymptotic solutions lead to the functional relations including the Q-function such as  Baxter's T-Q relation, which is left for future work. Since we take the bosonic reduction, the essential part of the solution is determined by the bosonic Lax operator. We will focus on the linear problem for the bosonic Lax operator.

\setcounter{equation}{0}	
\section{Modified $\mathcal{N}=1$ super sinh-Gordon equation}\label{Section 3}
We begin with studying the supersymmetric affine Toda field equations associated with 
$C(2)^{(2)}=osp(2|2)^{(2)}$. The Lie superalgebra $osp(2|2)$ has rank two and includes $sp(2)\oplus u(1)$ as a bosonic subalgebra. There are two simple fermionic roots $\beta_1=\delta_1-e_1$, $\beta_2=e_1+\delta_1$. The twisted affine Lie algebra is defined by ${\mathbb Z}_2$-automorphism $e_1\rightarrow -e_1$, where two simple roots are identified by this automorphism. The extended root is the lowest weight $\alpha_0=-1/2(\beta_{1}+\beta_{2})=-\delta_1$ in the odd generators. The generators of $osp(2|2)^{(2)}$ are given by
\begin{align}
	\begin{split}
			&E_{\alpha_1}=\frac{1}{\sqrt{2}}(E_{\beta_{1}}+E_{\beta_{2}}),\quad E_{\alpha_{0}}=\frac{1}{\sqrt{2}}(E_{-\beta_{1}}-E_{-\beta_{2}}),\\
			&E_{-\alpha_1}=\frac{1}{\sqrt{2}}(E_{-\beta_{1}}+E_{-\beta_{2}}),\quad E_{-\alpha_{0}}=\frac{1}{\sqrt{2}}(E_{\beta_{1}}-E_{\beta_{2}}).\\
	\end{split}
\end{align}
Then the modified affine Toda field equation \eqref{eq:modifiedaftfeq} becomes
\begin{align}
    D\bar{D}\Phi+
    \frac{m^2}{\beta}\alpha_{1}\exp(\beta \alpha_{1}\cdot\Phi)+\frac{m^2}{\beta}p(z)\bar{p}(\bar{z})\alpha_{0}\exp(\beta \alpha_{0}\cdot\Phi)=0,
\end{align}
which is nothing but the (modified) ${\cal N}=1$ sinh-Gordon equation \cite{Kulish:2005qc}.
The related linear problem and its bosonic reduction can be written down by the matrix representation of the generators.
There are two useful representations: 3-dimensional atypical representation and 4-dimensional typical representation \cite{Scheunert:1976wj,Marcu:1979se,Gotz:2005jz}. We will construct the asymptotic solutions to the linear problem.
	
\subsubsection*{4d representation}
It is straightforward to derive the bosonic linear problem in the case of four-dimensional representation. So we begin with this representation.
The generators are given by
\begin{align}
    E_{\alpha_{1}}&= \frac{1}{\sqrt{2}}(E_{1,4}+E_{2,4}+E_{3,1}+E_{3,2}), \quad E_{\alpha_{0}}=\frac{1}{\sqrt{2}}(-E_{1,3}+E_{2,3}-E_{4,1}+E_{4,2}), \nonumber\\
    E_{-\alpha_{1}}&=\frac{1}{\sqrt{2}}(-E_{1,3}-E_{2,3}+E_{4,1}+E_{4,2}), \quad E_{-\alpha_{0}}=\frac{1}{\sqrt{2}}(-E_{1,4}+E_{2,4}+E_{3,1}-E_{3,2}) \nonumber,\\
    \alpha_{1}\cdot H&=-\alpha_{0}\cdot H=\{ E_{\alpha_{1}},E_{-\alpha_{1}}\}=\{ E_{\alpha_{0}},E_{-\alpha_{0}}\}=-E_{3,3}+E_{4,4}.
    \label{eq:osp224d}
\end{align}
Here $E_{a,b}$ denotes the matrix with entries $(E_{a,b})_{ij}=\delta_{ai}\delta_{bj}$, 
whose size is equal to the dimension of the representation.
The super linear problem \eqref{eq:lax3} in $osp(2|2)^{(2)}$ becomes
 \begin{align}\label{eq:c2lax3}
     {\cal L}\Psi=\left\{ {\cal D}-\tilde{\theta}\frac{l_{1}\alpha_{1}\cdot H}{x} -\left( E_{\alpha_1} +p(x,E) E_{\alpha_0} \right)\right\}\Psi=0.
 \end{align}
Substituting the representation (\ref{eq:osp224d}) into (\ref{eq:c2lax3}), one obtains a basis of the asymptotic solution \eqref{eq:asymptoticinfity}
at infinity:
\begin{align}\label{eq:c2asymptoticinfity}
\begin{split}
    &\psi_{1}(x,\tilde{\theta})\sim\begin{pmatrix}
			-i\sqrt{2}x^{2M}\exp(-\frac{x^{2M+1}}{2M+1})\tilde{\theta}\\
			0\\
			-x^{-M}\exp(-\frac{x^{2M+1}}{2M+1})\\
			x^{M}\exp(-\frac{x^{2M+1}}{2M+1})\\
		\end{pmatrix},\quad \psi_{2}(x,\tilde{\theta})\sim\begin{pmatrix}
			\exp(-\frac{x^{2M+1}}{2M+1})\\
			0\\
			\frac{-ix^{M}}{\sqrt{2}}\exp(-\frac{x^{2M+1}}{2M+1})\tilde{\theta}\\
			\frac{ix^{3M}}{\sqrt{2}}\exp(-\frac{x^{2M+1}}{2M+1})\tilde{\theta}\\
		\end{pmatrix},\\
	&\psi_{3}(x,\tilde{\theta})\sim\begin{pmatrix}
			0\\
			-i\sqrt{2}x^{2M}\exp(\frac{x^{2M+1}}{2M+1})\tilde{\theta}\\
			x^{-M}\exp(\frac{x^{2M+1}}{2M+1})\\
			x^{M}\exp(\frac{x^{2M+1}}{2M+1})\\
		\end{pmatrix},\quad \psi_{4}(x,\tilde{\theta})\sim\begin{pmatrix}
			0\\
			\exp(\frac{x^{2M+1}}{2M+1})\\
			\frac{-ix^{M}}{\sqrt{2}}\exp(\frac{x^{2M+1}}{2M+1})\tilde{\theta}\\
			\frac{-ix^{3M}}{\sqrt{2}}\exp(\frac{x^{2M+1}}{2M+1})\tilde{\theta}\\
		\end{pmatrix},\\
		\end{split}
\end{align}
and the asymptotic solution around the origin \eqref{eq:asymptoticorigin}:
%in $osp(2,2)^{(2)}$ is of the form
    \begin{equation}
    	\chi_{1}(x,\tilde{\theta})\sim\begin{pmatrix}
			1\\
			0\\
			0\\
			0\\
		\end{pmatrix},\quad \chi_{2}(x,\tilde{\theta})\sim\begin{pmatrix}
			0\\
			1\\
			0\\
			0\\
		\end{pmatrix}, \quad \chi_{3}(x,\tilde{\theta})\sim\begin{pmatrix}
			0\\
			0\\
			x^{l}\\
			0\\
		\end{pmatrix}, \quad \chi_{4}(x,\tilde{\theta})\sim\begin{pmatrix}
			0\\
			0\\
			0\\
			x^{-l}\\
		\end{pmatrix}.
    \end{equation}
    
On the other hand, the bosonic linear problem \eqref{eq:bosoniclpcl1} for 4-component vector $\Psi_{0}$ lead to the Schr\"odinger equation with quadratic potential.
\begin{align}
		\begin{split}
		\mathcal{L}_{B}\Psi_{0}
		=&\begin{pmatrix}
			\partial_{x}+p(x,E) & 0 & 0 & 0 \\
			0 & \partial_{x}- p(x,E) & 0 & 0 \\
			0 & 0 & \partial_{x}+\frac{l}{x} & -1 \\
			0 & 0 & -p^{2}(x,E) & \partial_{x}-\frac{l}{x} \\
		\end{pmatrix}\begin{pmatrix}
			\Psi_{0}^{1}\\
			\Psi_{0}^{2}\\
			\Psi_{0}^{3}\\
			\Psi_{0}^{4}\\
		\end{pmatrix}=0.\\
	\end{split}
\end{align}
The equations for $\Psi_{0}^{1}$ and $\Psi_{0}^{2}$ become the first order differential equation:
\begin{align}
    (\partial_x- p(x,E))\Psi_{0}^{1}&=0,\quad
    (\partial_x+ p(x,E))\Psi_{0}^{2}=0.
\end{align}
It is easy to solve them
\begin{align}
    \Psi_{0}^{1}(x)&=\exp\left(-\int dx \;p(x,E) \right),\quad
   \Psi_{0}^{2}(x)=\exp\left( \int dx\; p(x,E) \right).
\end{align}
These solutions correspond to the $u(1)=o(2)$ subalgebra.
The third  and the fourth components, which correspond to the $sp(2)$ subalgebra, %in our representation are bosonic part and 
lead to the linear differential equations. These equations becomes the ODE for $\Psi^3_0$:
	\begin{equation}\label{eq:quadraticpotential}
		[-\partial_{x}^{2}+\frac{l(l+1)}{x^{2}}+(x^{2M}-E)^{2}]\Psi_{0}^{3}=0.
	\end{equation}
This equation provides the same differential equation with the potential $P_K(x)=(x^{2M\over K}-E)^K$ with $K=2$ and the replacement $M\rightarrow 2M$, which has been argued to correspond to ${\cal N}=1$ supersymmetric coset models \cite{Dorey:2006an}. 

For completeness, we repeat the derivation of the Bethe ansatz equation following \cite{Dorey:1998pt,Bazhanov:1998wj}.
The asymptotic solutions at infinity are given by
\begin{equation}\label{eq:infinityasymptotic}
    \psi_{+}(x)\sim \frac{1}{\sqrt{2i}}x^{-M}\exp(\frac{x^{2M+1}}{2M+1}),\quad \psi_{-}(x)\sim -\frac{1}{\sqrt{2i}}x^{-M}\exp(-\frac{x^{2M+1}}{2M+1}),
\end{equation}
and at the origin
\begin{equation}\label{eq:originasymptotic}
    \chi_{+}(x)\sim x^{l+1},\quad \chi_{-}(x)\sim x^{-l}.
\end{equation}
The rotated solution $\Psi_{0[k]}^{3}(x,E,l):=\omega^{k/2}\Psi_{0}^{3}(\omega^{k}x,\omega^{2Mk}E,l)$ $(k\in {\mathbb Z})$ still satisfy \eqref{eq:quadraticpotential} for  $\omega=e^{\frac{\pi i}{2M+1}}$. 
Let us introduce
the Wronskian  $W[f(x),g(x)]=f(\partial_{x}g)-(\partial_{x}f)g$ of functions
$f(x)$ and $g(x)$.
Then $\psi_{\pm}$ and $\chi_{\pm}$ are normalized as
%is independent with $x$ if $f$ and $g$ satisfy \eqref{eq:quadraticpotential}. Then make use of the asymptotic solution at infinity \eqref{eq:infinityasymptotic}, one can show $W[\Psi_{0[1/2]}^{3},\Psi_{0[-1/2]}^{3}]=
$W[\psi_{\pm[1/2]},\psi_{\pm[-1/2]}]=1$ and $W[\chi_{+},\chi_{-}]=-(2l+1)$. 
%According to the definition of Q-function \eqref{eq:Qfunction}, 
We define the Q-functions by
$\psi_{-}=\sum_{i=\pm}Q_{i}(E,l)\chi_{i}(x)$.
Here $Q_{+}$ and $Q_-$ are the same as $Q_{3}$ and $Q_{4}$ in \eqref{eq:Qfunction}, respectively. Substituting these expansions into the Wronskian $W[\psi_{-[1/2]},\psi_{-[-1/2]}]$, one obtains the equation
\begin{equation}\label{eq:QQrelation}
\omega^{\frac{-2l-1}{2}}(2l+1)
%W[\chi_{-},\chi_{+}]
Q_{-[\frac{1}{2}]}Q_{+[-\frac{1}{2}]}-\omega^{\frac{2l+1}{2}}(2l+1)
%W[\chi_{+},\chi_{-}]
Q_{+[\frac{1}{2}]}Q_{-[-\frac{1}{2}]}=1.
\end{equation}
 %where the Wronskian  in the limit $x\rightarrow0$. 
 Now set $E_{0}$ to be the zero point of $Q_{+}(E,l)=0$. Evaluating \eqref{eq:QQrelation} at $E=\omega^{M}E_{0}$ and at $E=\omega^{-M}E_{0}$, %and dividing the two equations thus obtained, 
 we find the Bethe ansatz equation \cite{Dorey:2006an}:
\begin{equation}\label{eq:bae1}
    \omega^{2l+1}\frac{Q_{+[1]}(E_{0},l)}{Q_{+[-1]}(E_{0},l)}=-1.
\end{equation}
Note that the equation (\ref{eq:bae1}) takes the same form for general $P_K(x)$ but the $K$ dependence enters through $\omega$.
\subsubsection*{3d representation}
The ODE \eqref{eq:quadraticpotential} would be independent of the choice of the representations.
To confirm this, we consider the three-dimensional representation.
The generators are
\begin{align}
	\begin{split}
		&E_{\alpha_1}=E_{1,2}+E_{2,3},\;\;\;\;\;\;\;\;
		E_{-\alpha_1}=E_{3,2}-E_{2,1},\\
		&E_{\alpha_{0}}=-E_{2,1}-E_{3,2},\;\;\;\;\;\;\;\;
		E_{-\alpha_{0}}=E_{1,2}-E_{2,3},\\
		&\alpha_{1}\cdot H=-\alpha_{0}\cdot H=\{ E_{\alpha_{1}},E_{-\alpha_{1}}\}=\{ E_{\alpha_{0}},E_{-\alpha_{0}}\}=E_{3,3}-E_{1,1}.
	\end{split}
\end{align}
Based on the super linear problem \eqref{eq:c2lax3}, 
the asymptotic solutions  at infinity are given by the following basis
\begin{align}
    \begin{split}
    &\psi_{1}(x,\tilde{\theta})\sim\begin{pmatrix}
    -x^{-M}\exp(-\frac{x^{2M+1}}{2M+1})\\
    -2ix^{2M}\exp(-2\frac{x^{2M+1}}{2M+1})\tilde{\theta}\\
    x^{M}\exp(-\frac{x^{2M+1}}{2M+1})\\
    \end{pmatrix}, \quad
    \psi_{2}(x,\tilde{\theta})\sim\begin{pmatrix}
    \frac{-i}{\sqrt{2}}x^{M}\exp(-2\frac{x^{2M+1}}{2M+1})\tilde{\theta}\\
    \frac{1}{\sqrt{2}}\exp(-2\frac{x^{2M+1}}{2M+1})\\
    \frac{i}{\sqrt{2}}x^{3M}\exp(-2\frac{x^{2M+1}}{2M+1})\tilde{\theta}\\
    \end{pmatrix},\\ 
    &\psi_{3}(x,\tilde{\theta})\sim\begin{pmatrix}
    x^{-M}\exp(\frac{x^{2M+1}}{2M+1})\\
    -2ix^{2M}\tilde{\theta}\\
    x^{M}\exp(\frac{x^{2M+1}}{2M+1})\\
    \end{pmatrix}.\\
    \end{split}
\end{align}
The asymptotic solutions at the origin are
\begin{equation}
    	\chi_{1}(x,\tilde{\theta})\sim\begin{pmatrix}
			x^{-l}\\
			0\\
			0\\
		\end{pmatrix},\quad \chi_{2}(x,\tilde{\theta})\sim\begin{pmatrix}
			0\\
			1\\
			0\\
		\end{pmatrix}, \quad \chi_{3}(x,\tilde{\theta})\sim\begin{pmatrix}
			0\\
			0\\
			x^{l}\\
		\end{pmatrix}.
    \end{equation}
The bosonic linear problem, however, takes a different form. The bosonic Lax operator is given by
\begin{equation}
	\mathcal{L}_{B}=\begin{pmatrix}
		\partial_{x}+\frac{l}{x}+p(x,E) & 0 & -1\\
		0 & \partial_{x}+2p(x,E) & 0 &\\
		p^{2}(x,E) & 0 & \partial_{x}-\frac{l}{x}+p(x,E)\\
	\end{pmatrix}.
\end{equation}
The equation ${\cal L}_B\Psi=0$ for $\Psi=(\Psi^1_0,\Psi^2_0,\Psi^3_0)^T$ reduces to two ODEs for $\Psi^1_0$ and $\Psi^2_0$: 
\begin{align}
	\begin{split}
		&(\partial_{x}+\frac{l}{x}+p(x))(\partial_{z}-\frac{l}{x}+p(x))\Psi_{0}^{1}=p^{2}(x,E)\Psi_{0}^{1},\\
		&(\partial+2p(x,E))\Psi_{0}^{2}=0.\\
	\end{split}
\end{align}
Here the second equation corresponds to the $u(1)$ subalgebra.
For the first equation, 
setting $\psi_{1}=\exp(-\int p(x))\psi_{1}'$, one can show that  $\psi_{1}'$ satisfies 
\begin{equation}
	(-\partial_{x}^{2}+\frac{l(l+1)}{x^{2}}+(x^{2M}-E)^{2})\Psi_{0}^{1}=0.
\end{equation}
Finally we obtain the same equation with \eqref{eq:quadraticpotential} and the same asymptotic solutions with \eqref{eq:infinityasymptotic} and \eqref{eq:originasymptotic}. From the BAE \eqref{eq:bae1}, one can obtain the integrals of motion via the Non-linear integral equations. It was shown that the integrals of motion corresponds to the ones from $\mathcal{N}=1$ quantum sin(h)-Gordon model in \cite{Babenko:2017fmu}

So far, we have given the solutions to the super Lax equation \eqref{eq:lax3} with the $C(2)^{(2)}$ Lie superalgebraic structure. Moreover, we have shown that the bosonic Lax equation \eqref{eq:bosoniclpcl1} for the $C(2)^{(2)}$ algebra leads to an ODE with squared potential terms. In the next section, we generalize our analysis to other purely fermionic algebra. 

\section{Linear problem for modified $\mathcal{N}=1$ super affine Toda field equations}\label{Section 4}
In this section, we study the bosonic linear problems for other classical affine Lie superalgebras with purely odd simple root systems in Table \ref{tab:table_liesuperalgebras}.
The associated ODEs can be understood as a generalization of \eqref{eq:c2asymptoticinfity}. We first discuss the untwisted affine Lie  superalgebras $A(n|n)^{(1)}$, $B(n|n)^{(1)}$ and $D(n|n-1)^{(1)}$ and then the twisted affine Lie superalgebras $A(n|n)^{(2)}$ and $D(n|n)^{(2)}$.
We will write down the matrix representation for the generators $E_{\pm \alpha_i}$ $(i=0,1,\dots,r)$. 
The generators for the Cartan subalgebra are defined by (\ref{eq:superalgebra}),
where the normalization factors $\epsilon_i$ are listed in table \ref{tab:normalization}.
\begin{table}[tbh]
\begin{center}
\begin{tabular}{|c|c|c|c|c|c|c|}
\hline
 & $D(n|n-1)^{(1)}$ & $B(n|n)^{(1)}$ & $D(n|n)^{(2)}$ & $C(2)^{(2)}$ & $A(n|n)^{(1)}$ & $A(n|n)^{(2)}$\\
\hline
$\epsilon_{i}$ & $(-1)^{i}$ & $(-1)^{i}$ & $(-1)^{n+1}$ & 1 & $(-1)^{i}$ & $(-1)^{i}$\\
\hline
$\epsilon_{0}$ &$-1$ & $-1$ & $-1$ &$ -1$ & $1$ & $(-1)^{n+1}$ \\
\hline
\end{tabular}
\end{center}
\caption{List of the normalization factors $\epsilon_i$ and $\epsilon_0$ for the representations used in this paper.}
\label{tab:normalization}
\end{table}
We also introduce 
\begin{equation}
    D(\ell)=\partial_x-\frac{\ell}{x},
\end{equation}
for a constant $\ell$ to simplify the notation.
	
\subsection{$A(n|n)^{(1)}=sl(n+1|n+1)^{(1)}/\mathbf{I}_{2n+2}$}
Let us start with the untwisted affine Lie superalgebra $A(n|n)^{(1)}$.
Its $(2n+2)$-dimensional representation is defined by
\begin{align}\label{eq:anan1}
		\begin{split}
			&%E_{\alpha_{1}}=E_{1,n+2},\;\;\;\;\;\; E_{\alpha_{2}}=E_{n+2,2},\;\;\;\;\;\;
			E_{\alpha_{2i-1}}=E_{i,n+1+i},\quad E_{\alpha_{2i}}=E_{n+1+i,i+1}\;
			(i=1,\dots,n), \quad E_{\alpha_{0}}=E_{2n+2,1},\\
			&%E_{-\alpha_{1}}=(E_{\alpha_{1}})^{T},\;\;\;\;\; E_{-\alpha_{2}}=-(E_{\alpha_{2}})^{T}\;\;\;\;\;
			E_{-\alpha_{2i-1}}=-(E_{\alpha_{2i-1}})^{T},\quad E_{-\alpha_{2i}}=-(E_{\alpha_{2i}})^{T}\;
			(i=1,\dots,n),\quad E_{-\alpha_{0}}=-(E_{\alpha_{0}})^{T}.\\
		\end{split}
	\end{align}
Then the bosonic reduction of the linear problem \eqref{eq:bosonicLaxoperator} now takes the block-diagonal form:
\begin{align}
	\begin{split}
		\mathcal{L}_{B}=&\partial_{x}-\sum_{j=1}^{r} \frac{l_{j}\alpha_{j}\cdot H}{x}-(\sum_{i=1}^{r} E_{\alpha_{i}}+p(x,E)E_{\alpha_{0}})^{2}
		=\begin{pmatrix}
			\mathbf{A} & \mathbf{0}\\
			\mathbf{0} & \mathbf{B}\\
		\end{pmatrix}_{(2n+2) \times (2n+2)
		}\\
	\end{split}
\end{align}
with $r=2n$. The $(n+1)$-dimensional matrices $\mathbf{A}$ and $\mathbf{B}$ are given by
\begin{equation}
	\mathbf{A}=\begin{pmatrix}
			D(\tilde{l}_{1}) & -1 &  &  & & \\
			& D(\tilde{l}_{2}) & -1 &  & & \\
			& & \ddots & & & \\
			&  &  &  & D(\tilde{l}_{n}) &-1 \\
			-p(x,E)& & & &  &D(\tilde{l}_{n+1}) \\
	\end{pmatrix},
\end{equation}
	and
\begin{equation}
	\mathbf{B}=\begin{pmatrix}
			D(\tilde{l}_{n+2}) & -1 &  &  & & \\
			& D(\tilde{l}_{n+3}) & -1 &  & & \\
			& & \ddots & & & \\
			&  &  &  & D(\tilde{l}_{2n+1}) &-1 \\
			-p(x,E)& & & &  &D(\tilde{l}_{2n+2}) \\
	\end{pmatrix}.
\end{equation}
Here we defined
\begin{align}
\begin{split}
    &\tilde{l}_{i}=l_{2i-1}-l_{2i-2}, \quad 1\leq i\leq n+1,\quad l_{0}=0,\\
    &\tilde{l}_{n+1+i}=l_{2i-1}-l_{2i},\quad 1\leq i\leq n+1, \quad l_{2n+2}=0.\\
\end{split}
\end{align}
The bosonic linear problem ${\cal L}_B\psi=0$ for $\psi=(\psi_1,\dots, \psi_{2n+2})^T$ is made of two independent diagonal blocks.
Each sector provides %gives 
the $A_n^{(1)}$ type linear problem.
For the components $\psi_1$ and $\psi_{n+2}$, the two sets of the $(n+1)$-the order ODEs of the $A_{n}$-type
\cite{Dorey:2006an,Dorey:2007zx}
\begin{align}
	\begin{split}
		&D(\tilde{l}_{n+1})\dots D(\tilde{l}_{2})D(\tilde{l}_{1})\psi_{1}=p(x,E)\psi_{1},\\
		&D(\tilde{l}_{2n+2})\dots D(\tilde{l}_{n+3})D(\tilde{l}_{n+2})
		\psi_{n+2}=p(x,E)\psi_{n+2}.  \\
	\end{split}
\end{align}

\subsection{$B(n|n)^{(1)}=osp(2n+1|2n)^{(1)}$}
We next study the affine Lie superalgebra $B(n|n)^{(1)}=osp(2n+1|2n)^{(1)}$. The algebra contains the bosonic subalgebra $B_{n}\oplus C_n$. Their $(4n+1)$-dimensional matrix representation is found to be
\begin{align}
		\begin{split}
			&E_{\alpha_{0}}=E_{4n+1,1}+E_{2n+1,2n+2},\quad E_{-\alpha_{0}}=E_{1,4n+1}-E_{2n+2,2n+1},\\
			&E_{\alpha_{2i-1}}=E_{4n+2-i,2n+2-i}+E_{i,2n+1+i},\quad E_{-\alpha_{2i-1}}=E_{2n+2-i,4n+2-i}-E_{2n+1+i,i},\quad 1\leq i\leq n,\\
			&E_{\alpha_{2i}}=E_{2n+1-i,4n+2-i}-E_{2n+1+i,i+1},\quad E_{-\alpha_{2i}}=E_{4n+2-i,2n+1-i}-E_{i+1,2n+1+i},\quad 1\leq i\leq n-2.\\
			&E_{\alpha_{2n}}=E_{n+1,3n+2}-E_{3n+1,n+1},\quad E_{-\alpha_{2n}}=E_{n+1,3n+1}-E_{3n+2,n+1}.\\
		\end{split}
		\label{eq:bnn1}
\end{align}
Then the bosonic Lax operator takes the block-diagonal form:
\begin{align}
	\begin{split}
		\mathcal{L}_{B}=\begin{pmatrix}
			\mathbf{A} & \mathbf{0}\\
			\mathbf{0} & \mathbf{B}\\
		\end{pmatrix}_{4n+1\times 4n+1}\\
	\end{split}
\end{align}
with $(2n+1)$-dimensional matrix $\mathbf{A}$ and $2n$-dimensional matrix $\mathbf{B}$, which are defined by
\begin{equation}
\small
	\mathbf{A}=\begin{pmatrix}
		D(\tilde{l}_{1}) & 1 &  &  & & & & & &\\
		& D(\tilde{l}_{2}) & 1 &  & & & & & &\\
		& & \ddots & & & & & & &\\
		&  &  & D(\tilde{l}_{n}) & 1 &  & & & &\\
		&  &  &  & \partial &-1 & & & &\\
		& & & &  &D(-\tilde{l}_{n}) &-1 & & &\\
		& &  & & & & D(-\tilde{l}_{n-1}) & -1 & &\\
		& &  & & & & &\ddots & &\\
		-p(x,E)&  & & & & &  &  & D(-\tilde{l}_{2}) & -1\\
		&p(x,E) & & &  &  &  &  &  & D(-\tilde{l}_{1})\\
	\end{pmatrix}
\end{equation}
and
\begin{equation}
\small
	\mathbf{B}=\begin{pmatrix}
		D(\tilde{l}_{n+1}) & 1 &  &  & & & & & &\\
		& D(\tilde{l}_{n+2}) & 1 &  & & & & & &\\
		& & \ddots & & & & & & &\\
		&  &  & D(\tilde{l}_{2n}) & 1 & & & & &\\
		& & & & D(-\tilde{l}_{2n}) &-1 & & & &\\
		& &  & & &\ddots & & & &\\
		&  & & & & &  & D(-\tilde{l}_{n+2}) & & -1\\
		-2p(x,E)& & & & &  &  &  &  & D(-\tilde{l}_{n+1})\\
	\end{pmatrix}.
	\end{equation}
Here
\begin{align}
\begin{split}
&\tilde{l}_{i}=l_{2i-1}-l_{2i-2},\quad 1\leq i\leq n+1, \quad l_{0}=0,\\
    &\tilde{l}_{n+i}=l_{2i-1}-l_{2i}, \quad 1\leq i\leq n.\\
\end{split}
\end{align}
The linear problem ${\cal L}_B\psi=0$ for $\psi=(\psi_1,\dots, \psi_{4n+1})^T$ reduces to two ODEs:
\begin{align}
	\begin{split}
	&D(-\tilde{l}_{1})\cdots D(-\tilde{l}_{n})\partial D(\tilde{l}_{n})\cdots D(\tilde{l}_{1})\psi_{1}=2(-1)^{n}\sqrt{p(x,E)}\partial\sqrt{p(x,E)}\psi_{1},\\
		&D(-\tilde{l}_{n+1})\cdots D(-\tilde{l}_{2n})D(\tilde{l}_{2n})\cdots D(\tilde{l}_{n+2})D(\tilde{l}_{n+1})\psi_{2n+2}=2(-1)^{n}p(x,E)\psi_{2n+2}.\\
	\end{split}
\end{align}
It can be seen that the first one is $B_n^{(1)}$ type and the second one is $C_n^{(1)}$ type in \cite{Ito:2013aea}.

\subsection{$D(n|n-1)^{(1)}=osp(2n|2n-2)^{(1)}$}
Next we will discuss the untwisted affine Lie superalgebra $D(n|n-1)^{(1)}=osp(2n|2n-2)^{(1)}$. This algebra contains $D_n\oplus C_{n-1}$ as a bosonic subalgebra. The $(4n-2)$-dimensional representation is chosen to be
\begin{align}
		\begin{split}
			&E_{\alpha_{0}}=E_{4n-2,1}+E_{2n,2n+1},\quad E_{-\alpha_{0}}=E_{1,4n-2}-E_{2n+1,2n},\\
			&E_{\alpha_{2i-1}}=E_{4n-1-i,2n+1-i}+E_{i,2n+i},\quad E_{-\alpha_{2i-1}}=E_{2n+1-i,4n-1-i}-E_{2n+i,i},\quad 1\leq i\leq n-1,\\
			&E_{\alpha_{2i}}=E_{2n-i,4n-1-i}-E_{2n+i,i+1},\quad E_{-\alpha_{2i}}=E_{4n-1-i,2n-i}+E_{i+1,2n+i},\quad 1\leq i\leq n-2.\\
			&E_{\alpha_{2n-2}}=E_{n+1,3n}-E_{3n-1,n},\quad E_{-\alpha_{2n-2}}=E_{3n,n+1}+E_{n,3n-1},\\
			&E_{\alpha_{2n-1}}=E_{n,3n}-E_{3n-1,n+1},\quad E_{-\alpha_{2n-1}}=E_{3n,n}+E_{n+1,3n-1}.
		\end{split}
	\end{align}
For $n\geq 2$,
the bosonic Lax operator \eqref{eq:bosoniclpcl1} 
takes the form
\begin{align}
		\begin{split}
			\mathcal{L}_{B}=\begin{pmatrix}
				\mathbf{A} & \mathbf{0}\\
				\mathbf{0} & \mathbf{B}\\
			\end{pmatrix}_{4n-2\times 4n-2},\\
		\end{split}
\end{align}
where  $\mathbf{A}$ and $\mathbf{B}$ are $2n$-dimensional and $(2n-2)$-dimensional matrices, respectively and given by
\begin{equation}
\small
		\mathbf{A}=\begin{pmatrix}
			D(\tilde{l}_{1}) & 1 &  &  & & & & & &\\
			& D(\tilde{l}_{2}) & 1 &  & & & & & &\\
			& & \ddots & & & & & & &\\
			&  &  & D(\tilde{l}_{n-1}) & 1 & 1 & & & &\\
			&  &  &  & D(\tilde{l}_{n}) & &-1 & & &\\
			& & & &  &D(-\tilde{l}_{n}) &-1 & & &\\
			& &  & & & & D(-\tilde{l}_{n-1}) & -1 & &\\
			& &  & & & & &\ddots & &\\
			-p(x,E)&  & & & & &  &  & D(-\tilde{l}_{2}) & -1\\
			&p(x,E) & & &  &  &  &  &  & D(-\tilde{l}_{1})\\
		\end{pmatrix}
\end{equation}
and
\begin{equation}
\small
		\mathbf{B}=\begin{pmatrix}
			D(\tilde{l}_{n+1}) & 1 &  &  & & & & & &\\
			& D(\tilde{l}_{n+2}) & 1 &  & & & & & &\\
			& & \dots & & & & & & &\\
			&  &  & D(\tilde{l}_{2n-1}) & 2 & & & & &\\
			& & & & D(-\tilde{l}_{2n-1}) &-1 & & & &\\
			& &  & & &\dots & & & &\\
			&  & & & & &  & D(-\tilde{l}_{n+2}) & & -1\\
			-2p(x,E)& & & & &  &  &  &  & D(-\tilde{l}_{n+1})\\
		\end{pmatrix}.
\end{equation}	
Here $\tilde{l}$'s are defined by
\begin{align}
\begin{split}
    &\tilde{l}_{i}=l_{2i-1}-l_{2i-2},\quad 1\leq i\leq n+1, \quad l_{0}=0,\\
    &\tilde{l}_{n+i}=l_{2i-1}-l_{2i}, \quad 1\leq i\leq n-2,\quad \tilde{l}_{n-1}=l_{2n-3}-l_{2n-2}+l_{2n-1}.\\
\end{split}
\end{align}
We obtain two ODEs for $\psi_1$ and $\psi_{2n+1}$:
\begin{align}
		\begin{split}
&D(-\tilde{l}_{1})\cdots D(-\tilde{l}_{n-1})\partial_{x}^{-1}D(\tilde{l}_{n})\cdots D(\tilde{l}_{1})\psi_{1}=4(-1)^{n-1}\sqrt{p(x,E)}\partial_{x}\sqrt{p(x,E)}\psi_{1},\\
&D(-\tilde{l}_{n+1})\cdots D(-\tilde{l}_{2n-1})D(l_{2n-1})\cdots D(\tilde{l}_{n+2})D(\tilde{l}_{n+1})\psi_{2n+1}=4(-1)^{n-1}p(x,E)\psi_{2n+1}.\\
	\end{split}
\end{align}
It can be seen that the first one is $D_{n}^{(1)}$ type and the second one is $C_{n-1}^{(1)}$ type in \cite{Ito:2013aea}.

Let us consider the $n=2$ case, where the Dynkin diagram takes the specific form as seen from Table \ref{tab:table_liesuperalgebras}.
The representation is given by
\begin{align}
		\begin{split}
			E_{\alpha_{0}}&=E_{6,3}+E_{2,5},\quad E_{-\alpha_{0}}=E_{3,6}-E_{5,2},\quad
			E_{\alpha_{1}}=E_{6,2}+E_{3,5},\quad E_{-\alpha_{1}}=E_{2,6}-E_{3,5},\\
			E_{\alpha_{2}}&=E_{4,6}-E_{5,1},\quad E_{-\alpha_{2}}=E_{6,4}+E_{1,5},\quad
			E_{\alpha_{3}}=E_{1,6}-E_{5,4},\quad E_{-\alpha_{3}}=E_{6,1}+E_{4,5}.\\
		\end{split}
	\end{align}
The bosonic Lax operator \eqref{eq:bosoniclpcl1} now becomes
\begin{align}
	\begin{split}
		\mathcal{L}_{B}=\begin{pmatrix}
			\partial_{x}+\frac{l_{2}-l_{3}}{x} & -1 & -p(x,E) & 0 & 0 & 0\\
			p(x,E) & \partial_{x}+\frac{l_{1}}{x} & 0 & p(x,E) & 0 & 0\\
			1 & 0 & \partial_{x}-\frac{l_{1}}{x} & 1 & 0 & 0\\
			0 & -1 & -p(x,E) & \partial_{x}+\frac{l_{3}-l_{2}}{x} & 0 & 0\\
			0 & 0 & 0 & 0 & \partial_{x}+\frac{l_{2}+l_{3}}{x} & 2\\
			0 & 0 & 0 & 0 & -2p(x,E) & \partial_{x}-\frac{l_{2}+l_{3}}{x}\\
		\end{pmatrix}.
	\end{split}
\end{align}
The bosonic linear problem ${\cal L}_B\psi=0$ for $\psi=(\psi_1,\dots, \psi_6)^T$ leads to only one equation corresponding to $A_{1}^{(1)}$ type:
\begin{equation}
    (\partial_{x}-\frac{l_{2}+l_{3}}{x})(\partial_{x}+\frac{l_{2}+l_{3}}{x})\psi_{5}=-4p(x,E)\psi_{5}.
\end{equation}
Other components do not lead to a simple ODE.
We note that a similar thing also would happen for $D(2|1;\alpha)^{(1)}$, since for $\alpha=1$ the algebra becomes $D(2|1)$.

So far we have studied the untwisted affine Lie algebras with purely odd simple roots. We observed that the resulting ODEs are those of the bosonic subalgebras.

\subsection{$A(n|n)^{(2)}=sl(n+1|n+1)^{(2)}/\mathbf{I}_{2n+2}$}
Now we construct the bosonic Lax operator from the twisted affine Lie superalgebras. In this case, the generators are identified with 
${\mathbb Z}_2$-automorphism.
First we focus on the $A(n|n)^{(2)}$ type twisted affine Lie superalgebra,
which has $o(n+1)\oplus o(n+1)$ even subalgebra\cite{FSS}. We need to treat $n$ even and odd cases, separately. In fact, the Dynkin diagram of $A(n|n)^{(2)}$ for even $n$ is similar to $D(n|n)^{(2)}$, the diagram of $A(n|n)^{(2)}$ for odd $n$ is similar to $D(n|n-1)^{(1)}$.

\subsubsection{$A(2n|2n)^{(2)}$}
The algebra contains $B_n\oplus B_n$ as a bosonic subalgebra.
The representation is
\begin{align}
		\begin{split}
			&E_{\alpha_{2i}}=E_{n+1-i,4n+3-i}+E_{2n+1+i,n+1+i},\quad E_{-\alpha_{2i}}=E_{4n+3-i,n+1-i}-E_{n+1+i,2n+1+i},\quad i=1,\dots,n,\\
			&E_{\alpha_{2i+1}}=E_{n+1+i,2n+2+i}+E_{4n+2-i,n+1-i},\quad E_{-\alpha_{2i+1}}=E_{n+1-i,4n+2-i}-E_{2n+2+i,n+1+i},\quad i=0,\dots,n-1.\\
		\end{split}
	\end{align}
where we define $\alpha_{0}=\alpha_{n+1}$. The bosonic Lax operator \eqref{eq:bosoniclpcl1} %for $sl(2n+1,2n+1)^{(2)}$ algebra 
takes the form
\begin{align}
		\begin{split}
			\mathcal{L}_{B}=\begin{pmatrix}
				\mathbf{A} & \mathbf{0}\\
				\mathbf{0} & \mathbf{B}\\
			\end{pmatrix}_{4n+2\times 4n+2}\\
		\end{split}
\end{align}
with $(2n+1)$-dimensional matrices $\mathbf{A}$ and $\mathbf{B}$, which are defined by
\begin{equation}
\small
		\mathbf{A}=\begin{pmatrix}
			D(\tilde{l}_{1}) & -1 &  &  & & & & & &\\
			& D(\tilde{l}_{2}) & -p(x,E) &  & & & & & &\\
			&  & D(\tilde{l}_{3}) & -1 & & & & & &\\
			& & & \ddots & & & & & &\\
			&  &  &  & D(\tilde{l}_{n+1}) & -1 & & & &\\
			& & &  & & \ddots & & & &\\
			&  & & & &  & D(-\tilde{l}_{3}) &-p(x,E) & &\\
			&  & & & &  &  & D(-\tilde{l}_{2}) & -1 &\\
			-1 & & & &  & & & &D(-\tilde{l}_{1}) &\\
		\end{pmatrix}
\end{equation}
and
\begin{equation}
\small
		\mathbf{B}=\begin{pmatrix}
			D(\tilde{l}'_{1}) & -1 &  &  & & & & & &\\
			& D(\tilde{l}'_{2}) & -p(x,E) &  & & & & & &\\
			&  & D(\tilde{l}'_{3}) & -1 & & & & & &\\
			& & & \ddots & & & & & &\\
			&  &  &  & D(\tilde{l}'_{n+1}) & -1 & & & &\\
			& & &  & & \ddots & & & &\\
			&  & & & &  & D(-\tilde{l}'_{3}) &-p(x,E) & &\\
			&  & & & &  &  & D(-\tilde{l}'_{2}) & -1 &\\
			-1 & & & &  & & & &D(-\tilde{l}'_{1}) &\\
		\end{pmatrix},
\end{equation}
where
\begin{align}
\begin{split}
    &\tilde{l}_{i}=l_{2n-2i+3}-l_{2n-2i+2}, \quad 1\leq i\leq n, \quad l_{n+1}=0,\quad \tilde{l}_{n+1}=0\\
    &\tilde{l}'_{i}=l_{2i-1}-l_{2i},\quad 1\leq i\leq n, \quad l_{n+1}=0,\quad \tilde{l}'_{n+1}=0\\
\end{split}
\end{align}
The bosonic linear problem ${\cal L}_B\psi=0$ for $\psi=(\psi_1,\dots,\psi_{4n+2})^T$ takes the different form of that of the untwisted one $A(2n|2n)^{(1)}$ because their extended simple roots are different.
The ODE for $\psi_1$, which corresponds to the block $\textbf{A}$, is given as
\begin{equation}
		D(-\tilde{l}_{1}) D(-\tilde{l}_{2})p^{-1}(x,E)D(-\tilde{l}_{3})\cdots D(\tilde{l}_{n+1})\cdots D(\tilde{l}_{3})p^{-1}(x,E)D(\tilde{l}_{2})D(\tilde{l}_{1})\psi_{1}=\psi_{1}.
\end{equation}
The block matrix $\textbf{B}$ provides also the $A_{2n}$-type ordinary differential equation with different $\tilde{l}$'s.
\subsubsection{$A(2n-1|2n-1)^{(2)}$}
We next consider the twisted affine Lie algebra $A(2n-1|2n-1)^{(2)}$, whose bosonic subalgebra is $D_n\oplus D_n$.
The $4n$-dimensional representation is given by
\begin{align}
		\begin{split}
			&E_{\alpha_{2i}}=E_{n+1-i,4n+1-i}+E_{2n+i,n+i},\quad E_{-\alpha_{2i}}=E_{4n+1-i,n+1-i}-E_{n+i,2n+i},\quad 1\leq i \leq n-1,\\
			&E_{\alpha_{2i+1}}=E_{n+i,2n+1+i}+E_{4n-i,n+1-i},\quad E_{-\alpha_{2i+1}}=E_{n+1-i,4n-i}-E_{2n+1+i,n+i},\quad 0\leq i \leq n.\\
		\end{split}
	\end{align}
where $\alpha_{0}=\alpha_{n+1}$. The bosonic Lax operator \eqref{eq:bosoniclpcl1} %for $sl(2n,2n)^{(2)}$ algebra 
is of the form
\begin{align}
		\begin{split}
			\mathcal{L}_{B}=\begin{pmatrix}
				\mathbf{A} & \mathbf{0}\\
				\mathbf{0} & \mathbf{B}\\
			\end{pmatrix}_{4n\times 4n}\\
		\end{split}
\end{align}
with $2n$-dimensional matrices $\mathbf{A}$ and $\mathbf{B}$, which are defined by
\begin{equation}
\small
		\mathbf{A}=\begin{pmatrix}
			D(\tilde{l}_{1}) & -1 &  &  & & & & & &\\
			& D(\tilde{l}_{2}) & -p(x,E) &  & & & & & &\\
			&  & D(\tilde{l}_{3}) & -1 & & & & & &\\
			& & & \ddots & & & & & &\\
			&  &  &  & D(\tilde{l}_{n}) & -2 & & & &\\
			&  &  &  &  & D(-\tilde{l}_{n}) & -1 & & &\\
			& & &  & &  &\ddots & & &\\
			&  & & & &  &  &D(-\tilde{l}_{3}) & p(x,E) &\\
			-1 &  & & & &  &  &  & D(-\tilde{l}_{2}) & -1\\
			& -1 & & &  & & & & & D(-\tilde{l}_{1})\\
		\end{pmatrix}
\end{equation}
and
\begin{equation}
\small
		\mathbf{B}=\begin{pmatrix}
			D(\tilde{l}'_{1}) & -1 &  &  & & & & & &\\
			& D(\tilde{l}'_{2}) & -p(x,E) &  & & & & & &\\
			&  & D(\tilde{l}'_{3}) & -1 & & & & & &\\
			& & & \ddots & & & & & &\\
			&  &  &  & D(\tilde{l}'_{n}) & -2 & & & &\\
			&  &  &  &  & D(-\tilde{l}'_{n}) & -1 & & &\\
			& & &  & &  &\ddots & & &\\
			&  & & & &  &  &D(-\tilde{l}'_{3}) & p(x,E) &\\
			-1&  & & & &  &  &  & D(-\tilde{l}'_{2}) & -1\\
			& -1 & & &  & & & & & D(-\tilde{l}'_{1})\\
		\end{pmatrix}.
\end{equation}
Here
\begin{align}
\begin{split}
    &\tilde{l}_{i}=l_{2n-2i+3}-l_{2n-2i+2}, \quad 1\leq i\leq n-1,\quad \tilde{l}_{n}=l_{2}+l_{1}-l_{3},\quad l_{n+1}=0,\\
    &\tilde{l}'_{i}=l_{2i-1}-l_{2i},\quad 1\leq i\leq n-1, \quad \tilde{l}'_{n}=l_{4n-1}-l_{4n}-l_{4n+1},\quad l_{n+1}=0.\\
\end{split}
\end{align}
The bosonic linear problem ${\cal L}_B\psi=0$ for $\psi=(\psi_1,\dots,\psi_{4n})^T$ reduces to two set of the linear problem.
The corresponding ODE for the component $\psi_1$ is given as follows
\begin{equation}
		D(-\tilde{l}_{1}) D(-\tilde{l}_{2})\; p^{-1}(x,E)D(-\tilde{l}_{3})\cdots D(-\tilde{l}_{n})D(\tilde{l}_{n})\cdots D(\tilde{l}_{3})\; p^{-1}(x,E)D(\tilde{l}_{2})D(\tilde{l}_{1})\psi_{1}=2\partial\psi_{1}.
\end{equation}
The block matrix $\textbf{B}$ gives the same type of ordinary differential equation with different $\tilde{l}$'s.

The Dynkin diagram for $A(1|1)^{(2)}$ takes a different form of $A(2n-1|2n-1)^{(2)}$ ($n\geq 2$). The representation is chosen to be
\begin{align}
		\begin{split}
			E_{\alpha_{0}}&=E_{3,2}+E_{1,4},\quad 
			E_{\alpha_{1}}=E_{4,2}+E_{1,3},\quad 
			E_{\alpha_{2}}=E_{3,1}+E_{2,4},\\
			E_{-\alpha_{0}}&=E_{2,3}-E_{4,1},\quad
			E_{-\alpha_{1}}=E_{2,4}-E_{3,1},\quad
			E_{-\alpha_{2}}=E_{1,3}-E_{4,2}.\\
		\end{split}
	\end{align}
Then the linear problem takes a specific form.
%others 
%and it can not be generalized to higher dimensions. 
The bosonic Lax operator \eqref{eq:bosoniclpcl1} %now 
becomes
\begin{align}
		\begin{split}
			\mathcal{L}_{B}=\left(
			\begin{array}{cccc}
				\partial_{x}+\frac{l_{1}-l_{2}}{x} & -2  p(x,E) & 0 & 0 \\
				0 & \partial_{x}-\frac{l_{1}-l_{2}}{x} & 0 & 0 \\
				0 & 0 & \partial_{x}+\frac{l_{1}-l_{2}}{x} & -2  p(x,E) \\
				0 & 0 & 0 & \partial_{x}-\frac{l_{1}-l_{2}}{x} \\
			\end{array}
			\right)-\mathbf{I},
		\end{split}
\end{align}
where the identity matrix $\textbf{I}$ can be modded out in $A(1|1)^{(2)}=sl(2|2)^{(2)}/\mathbf{I}$. The bosonic linear problem ${\cal L}_B\psi=0$ for $\psi=(\psi_1,\dots, \psi_4)^T$ leads to the set of equations:
\begin{align}
		\begin{split}
			&(\partial_{x}+\frac{l_{1}-l_{2}}{x})\psi_{1}=2 p(x,E)\psi_{2},\\
			&(\partial_{x}-\frac{l_{1}-l_{2}}{x})\psi_{2}=0.\\
		\end{split}
\end{align}
so the ODE becomes
\begin{equation}
		(\partial_{x}-\frac{l_{1}-l_{2}}{x})p(x,E)^{-1}(\partial_{x}+\frac{l_{1}-l_{2}}{x})\psi_{1}=0.
\end{equation}
where $\psi_{3}$ satisfies the same equation.
\subsection{$D(n|n)^{(2)}=osp(2n|2n)^{(2)}$}
Finally, we study the twisted affine Lie superalgebra $D(n|n)^{(2)}=osp(2n|2n)^{(2)}$.
The $n=1$ case of $osp(2|2)^{(2)}$ has been discussed in Section \ref{Section 3}. The algebra $D(n|n)^{(2)}$ contains the bosonic subalgebra $B_{n-1}\oplus C_n$.
The $(4n-1)$-dimensional representation is chosen to be
\begin{align}
		\begin{split}
			&E_{\alpha_{0}}=E_{n,2n+1}+E_{4n,n},\quad E_{-\alpha_{0}}=E_{2n+1,n}-E_{n,4n},\\
			&E_{\alpha_{2i-1}}=E_{2n+1-i,4n+1-i}-E_{2n+i,i},\quad E_{-\alpha_{2i-1}}=E_{4n+1-i,2n+1-i}+E_{i,2n+i},\quad 1\leq i\leq n-1,\\
			&E_{\alpha_{2i}}=E_{4n-i,2n+1-i}+E_{i,2n+1+i},\quad E_{-\alpha_{2i}}=E_{2n+1-i,4n-i}-E_{2n+1+i,i},\quad 1\leq i\leq n-1,\\
			&E_{\alpha_{2n-1}}=E_{n+1,3n+1}-E_{3n,n+1},\quad E_{-\alpha_{2n-1}}=E_{3n+1,n+1}+E_{n+1,3n}.\\
		\end{split}
	\end{align}
The bosonic Lax operator \eqref{eq:bosoniclpcl1} now becomes
\begin{align}
		\begin{split}
			\mathcal{L}_{B}=\begin{pmatrix}
				\mathbf{A} & \mathbf{0}\\
				\mathbf{0} & \mathbf{B}\\
			\end{pmatrix}_{4n\times 4n}\\
		\end{split}
\end{align}
with $\mathbf{A}$ and $\mathbf{B}$ being $2n$-dimensional and $2n$-dimensional matrices, respectively. They are given by
\begin{equation}
		\mathbf{A}=\begin{pmatrix}
			D(\tilde{l}_{1}) & 1 &  &  & & & & & &\\
			& D(\tilde{l}_{2}) & 1 &  & & & & & &\\
			& & \ddots & & & & & & &\\
			&  &  & D(\tilde{l}_{n-1}) & 0 & 1 & & & &\\
			p(x,E)&  &  &  & \partial & & & & &\\
			& & & &  &\partial &-1 & & &\\
			& &  & & & & D(-\tilde{l}_{n-1}) & -1 & &\\
			& &  & & & & &\ddots & &\\
			&  & & & & &  &  & D(-\tilde{l}_{2}) & -1\\
			& & & & -p(x,E) &  &  &  &  & D(-\tilde{l}_{1})\\
		\end{pmatrix}
\end{equation}
and
\begin{equation}
		\mathbf{B}=\begin{pmatrix}
			D(\tilde{l}_{n+1}) & 1 &  &  & & & & & &\\
			& D(\tilde{l}_{n+2}) & 1 &  & & & & & &\\
			& & \dots & & & & & & &\\
			&  &  & D(\tilde{l}_{2n}) & 1 & & & & &\\
			& & & & D(-\tilde{l}_{2n}) &-1 & & & &\\
			& &  & & &\dots & & & &\\
			&  & & & & &  & D(-\tilde{l}_{n+2}) & & -1\\
			-p^{2}(x,E)& & & & &  &  &  &  & D(-\tilde{l}_{n+1})\\
		\end{pmatrix}
\end{equation}
where
\begin{align}
\begin{split}
    &\tilde{l}_{i}=l_{2i}-l_{2i-1},\quad 1\leq i\leq n-1,\\
    &\tilde{l}_{n+i}=l_{2i-2}-l_{2i-1}, \quad 1\leq i\leq n,\quad l_{0}=0.\\
\end{split}
\end{align}
The bosonic linear problem ${\cal L}_B\psi=0$  for $\psi=(\psi_1,\dots, \psi_{4n-1})^T$ splits into two set of equations.
The first one provides the $D_{n}^{(2)}$ type ODE in \cite{Ito:2013aea} (or $C_{n-1}$-type ODE in \cite{Dorey:2006an}): %is a new type
\begin{equation}
		D(-\tilde{l}_{1})D(-\tilde{l}_{2})\cdots D(-\tilde{l}_{n-1})\partial_{x} D(\tilde{l}_{n-1})\cdots D(\tilde{l}_{1})\psi_{1}=(-1)^{n}p(x,E)\partial^{-1}_{x}p(x,E)\psi_{1}.
\end{equation}
The BAE corresponding to this ODE has been studied in \cite{Dorey:2006an}.
The second one gives the ordinary differential equation of $C^{(1)}_{n}$ type with $p(z)\rightarrow p^{2}(z)$:
%which seems to be the generalization of $C^{(1)}_{n}$ algebra with $p(z)\rightarrow p^{2}(z)$:
\begin{equation}
		D(-\tilde{l}_{n+1})D(-\tilde{l}_{n+2})\cdots D(-\tilde{l}_{2n})D(\tilde{l}_{2n})\cdots D(\tilde{l}_{n+2})D(\tilde{l}_{n+1})\psi_{2n+1}=(-1)^{n}p^{2}(x,E)\psi_{2n+1}.
		\label{eq:odedndn2}
\end{equation}
This is a generalization of the ODE for $osp(2|2)^{(2)}$ to higher order ones.
The ODE is regarded as the $A_{2n-1}^{(1)}$ type ODE with ${\mathbb Z}_2$-symmetric $\tilde{l}$ \cite{Ito:2020htm} and the squared potential $p^2$. The associated BAE can be obtained from those of $A_{2n-1}^{(1)}$-type by imposing ${\mathbb Z}_2$-symmetry. The squared potential implies the correspondence between the ODE and the coset model $SU(2n)_2 \times SU(2n)_L/SU(2n)_{2+L}$ which can be viewed as a supersymmetric theory \cite{Dorey:2006an}.
\begin{table}[h]\label{table5}
		\small
		\renewcommand{\arraystretch}{1.5}
		\begin{tabular}{|c|c|}
			\hline
			$A(n|n)^{(1)}$ & $A_n^{(1)}:\;D(\tilde{l}_{n+1})\dots D(\tilde{l}_{2})D(\tilde{l}_{1})\psi_{1}=p\psi_{1}$ \\ 
			& $A_n^{(1)}:\; D(\tilde{l}_{2n+2})\dots D(\tilde{l}_{n+3})D(\tilde{l}_{n+2})
		\psi_{n+2}=p\psi_{n+1}$\\ \hline
			$D(n|n-1)^{(1)}$ & $C_{n-1}^{(1)}:\; D(-\tilde{l}_{1})\cdots D(-\tilde{l}_{n-1})D(l_{n-1})\cdots D(\tilde{l}_{2})D(\tilde{l}_{1})\psi_{1}=4p\psi_{1}$ \\
			& $D_n^{(1)}:\; D(-\tilde{l}_{n})\cdots D(-\tilde{l}_{2n-2})\partial_{x}^{-1}D(\tilde{l}_{2n-1})\cdots D(\tilde{l}_{n})\psi_{n}=4\sqrt{p}\partial_{x}\sqrt{p}\psi_{n}$\\ \hline
			$B(n|n)^{(1)}$ & $C_n^{(1)}:\;D(-\tilde{l}_{1})\cdots D(-\tilde{l}_{n})D(\tilde{l}_{n})\cdots D(\tilde{l}_{2})D(\tilde{l}_{1})\psi_{1}=2p\psi_{1}$\\ 
			& $B_n^{(1)}:\;D(-\tilde{l}_{n+1})\cdots D(-\tilde{l}_{2n})\partial D(\tilde{l}_{2n-1})\cdots D(\tilde{l}_{n+1})\psi_{n+1}=2\sqrt{p}\partial\sqrt{p}\psi_{n+1}$\\ \hline
			$C(2)^{(2)}$ & $D(-l)D(l)\psi=p^{2}\psi$\\ \hline
			$A(2n|2n)^{(2)}$ & $D(-\tilde{l}_{1}) D(-\tilde{l}_{2})p^{-1}D(-\tilde{l}_{3})\cdots D(\tilde{l}_{n+1})\cdots D(\tilde{l}_{3})p^{-1}D(\tilde{l}_{2})D(\tilde{l}_{1})\psi_{1}=\psi_{1}$\\\hline
			$A(2n-1|2n-1)^{(2)}$ & $D(-l_{1}) D(-l_{2})\; p^{-1}D(-l_{3})\cdots D(-l_{n})$\\
			&$\times D(l_{n})\cdots D(l_{3})\; p^{-1}D(l_{2})D(l_{1})\psi_{1}=2\partial\psi_{1}$\\ \hline
			$D(n|n)^{(2)}$ & $D(-\tilde{l}_{1})D(-\tilde{l}_{2})\dots D(-\tilde{l}_{n})D(\tilde{l}_{n})\dots D(\tilde{l}_{2})D(\tilde{l}_{1})\psi=-p^{2}\psi$\\ 
			& $D(-\tilde{l}_{n+1})D(-\tilde{l}_{n+2})\cdots D(-\tilde{l}_{2n-1})\partial_{x} D(\tilde{l}_{2n-1})\cdots D(\tilde{l}_{n+1})\psi_{1}=(-1)^{n}p\partial^{-1}_{x}p\psi_{1}$\\ \hline
		\end{tabular}
		\caption{Summary of ODEs for affine Lie superagebras}
\end{table}
\section{Conclusions and Discussion}
In this paper, we have studied the modified $\mathcal{N}=1$ supersymmetric affine Toda field equations and their associated linear problem for classical affine Lie superalgebras with purely fermionic simple root systems. First, from $\mathcal{N}=1$ super affine Toda field equations, we have found the super Lax equations and formulated  their bosonic linear problem.
We also discussed the conformal limit of the linear problem, which takes the form of  the first-order differential equations in ${\cal N}=1$ superspace.
In particular, we consider the bosonic reduction of the linear problem.  
In some cases, it reduces to two higher-order ODEs related to the bosonic subalgebra of the affine Lie superalgebra. 
The ODEs are summarized in Table \ref{table5}.
For the untwisted affine Lie superalgebra, the ODEs appear in the context of the ODE/IM correspondence for affine Lie algebras.
For the twisted affine Lie algebra $C(2)^{(2)}$, one finds the second-order ODE, which is known to be related to the supersymmetric minimal models in the previous literature.
For other twisted affine Lie superalgebras, the associated ODEs are new.
Hence it is desirable to study the integrable systems using the ODE/IM correspondence. 
It is also interesting to study the ODE/IM correspondence for affine Lie superalgebra $D(2|1;\alpha)^{(1)}$ since this algebra contains three bosonic $sl(2)$ subalgebras. 
So far, the ODE/IM correspondence for $osp(2|2)^{(2)}$ superalgebra has been verified by comparing the integrals of motion on both sides \cite{Babenko:2017fmu}. We should mention both the IM side \cite{Kulish:2005qc} and the ODE side found that in $\mathcal{N}=1$ super sine(h)-Gordon model, it is the second-order transfer matrix that generates the integrals of motion. On the IM side, it results from the 3d atypical representation of $osp(2|2)^{(2)}$ superalgebra that is chosen in quantum $\mathcal{N}=1$ super sine(h)-Gordon model \cite{Babenko:2017fmu}, while on the ODE side, it is from the quadratic potential. It will be interesting to calculate the Thermodynamic Bethe ansatz equations and figure out the profound reason for this problem. On the other hand, it would be worthwhile to construct the quantum integrable models for other affine Lie superalgebras, which are known in the previous literatures \cite{Tsuboi:1997iq,Tsuboi:1999ft}. Finally, it is interesting to apply to the superstring theory on the AdS spacetime, which are realized as the coset model of the supergroup\cite{Bena:2003wd}.

\subsection*{Acknowledgements}
We would like to thank C. Locke and H. Shu for useful discussions.
The work of K.I. is supported in part by Grant- in-Aid for Scientific Research 21K03570, 18K03643 from Japan Society for the Promotion of Science (JSPS). The work of M. Z. is supported by Advanced Research Center for Quantum Physics and Nanoscience, Tokyo Institute of Technology.


\begin{thebibliography}{99}
\bibitem{Lukyanov:2010rn}
S.~L.~Lukyanov and A.~B.~Zamolodchikov,
%``Quantum Sine(h)-Gordon Model and Classical Integrable Equations,''
JHEP \textbf{07} (2010), 008
%doi:10.1007/JHEP07(2010)008
[arXiv:1003.5333 [math-ph]].

\bibitem{Dorey:2012bx}
P.~Dorey, S.~Faldella, S.~Negro and R.~Tateo,
%``The Bethe Ansatz and the Tzitzeica-Bullough-Dodd equation,''
Phil. Trans. Roy. Soc. Lond. A \textbf{371} (2013), 20120052
%doi:10.1098/rsta.2012.0052
[arXiv:1209.5517 [math-ph]].
		
\bibitem{Ito:2013aea}
K.~Ito and C.~Locke,
%``ODE/IM correspondence and modified affine Toda field equations,''
Nucl. Phys. B \textbf{885} (2014), 600-619
%doi:10.1016/j.nuclphysb.2014.06.007
[arXiv:1312.6759 [hep-th]].
		
\bibitem{Adamopoulou:2014fca}
P.~Adamopoulou and C.~Dunning,
%``Bethe Ansatz equations for the classical $A_n^{(1)}$ affine Toda field theories,''
J. Phys. A \textbf{47} (2014), 205205
%doi:10.1088/1751-8113/47/20/205205
[arXiv:1401.1187 [math-ph]].

\bibitem{Ito:2018wgj}
K.~Ito and H.~Shu,
%``Massive ODE/IM Correspondence and Non-linear Integral Equations for $A_r^{(1)}$-type modified Affine Toda Field Equations,''
J. Phys. A \textbf{51} (2018) no.38, 385401
%doi:10.1088/1751-8121/aad63f
[arXiv:1805.08062 [hep-th]].
	
\bibitem{Alday:2010vh}
L.~F.~Alday, J.~Maldacena, A.~Sever and P.~Vieira,
%``Y-system for Scattering Amplitudes,''
J. Phys. A \textbf{43} (2010), 485401
%doi:10.1088/1751-8113/43/48/485401
[arXiv:1002.2459 [hep-th]].

\bibitem{Hatsuda:2010cc}
Y.~Hatsuda, K.~Ito, K.~Sakai and Y.~Satoh,
%``Thermodynamic Bethe Ansatz Equations for Minimal Surfaces in $AdS_{3}$,''
JHEP \textbf{04} (2010), 108
%doi:10.1007/JHEP04(2010)108
[arXiv:1002.2941 [hep-th]].

\bibitem{Dorey:1998pt}
P.~Dorey and R.~Tateo,
%``Anharmonic oscillators, the thermodynamic Bethe ansatz, and nonlinear integral equations,''
J. Phys. A \textbf{32} (1999), L419-L425
%doi:10.1088/0305-4470/32/38/102
[arXiv:hep-th/9812211 [hep-th]].

\bibitem{Bazhanov:1998wj}
V.~V.~Bazhanov, S.~L.~Lukyanov and A.~B.~Zamolodchikov,
%``Spectral determinants for Schrodinger equation and Q operators of conformal field theory,''
J. Stat. Phys. \textbf{102} (2001), 567-576
%doi:10.1023/A:1004838616921
[arXiv:hep-th/9812247 [hep-th]].

\bibitem{Ito:2018eon}
K.~Ito, M.~Mari\~no and H.~Shu,
%``TBA equations and resurgent Quantum Mechanics,''
JHEP \textbf{01} (2019), 228
%doi:10.1007/JHEP01(2019)228
[arXiv:1811.04812 [hep-th]].

\bibitem{Ito:2021boh}
K.~Ito, T.~Kondo, K.~Kuroda and H.~Shu,
%``WKB periods for higher order ODE and TBA equations,''
JHEP \textbf{10} (2021), 167
%doi:10.1007/JHEP10(2021)167
[arXiv:2104.13680 [hep-th]].

\bibitem{Ito:2021sjo}
K.~Ito, T.~Kondo and H.~Shu,
%``Wall-crossing of TBA equations and WKB periods for the third order ODE,''
Nucl. Phys. B \textbf{979} (2022), 115788
%doi:10.1016/j.nuclphysb.2022.115788
[arXiv:2111.11047 [hep-th]].

\bibitem{Dorey:2006an}
P.~Dorey, C.~Dunning, D.~Masoero, J.~Suzuki and R.~Tateo,
%``Pseudo-differential equations, and the Bethe ansatz for the classical Lie algebras,''
Nucl. Phys. B \textbf{772} (2007), 249-289
%doi:10.1016/j.nuclphysb.2007.02.029
[arXiv:hep-th/0612298 [hep-th]].

\bibitem{Lukyanov:2006gv}
S.~L.~Lukyanov,
%``Notes on parafermionic QFT's with boundary interaction,''
Nucl. Phys. B \textbf{784} (2007), 151-201
%doi:10.1016/j.nuclphysb.2007.04.034
[arXiv:hep-th/0606155 [hep-th]].

\bibitem{Babenko:2017fmu}
C.~Babenko and F.~Smirnov,
%``Suzuki equations and integrals of motion for supersymmetric CFT,''
Nucl. Phys. B \textbf{924} (2017), 406-416
%doi:10.1016/j.nuclphysb.2017.09.019
[arXiv:1706.03349 [hep-th]].
%C. Babenko, F. Smirnov 2017 Suzuki equations and integrals of motion for supersymmetric CFT. Nuclear Physics B 924(C)

\bibitem{Suzuki:1998ve}
J.~Suzuki,
%``Spinons in magnetic chains of arbitrary spins at finite temperature,''
J. Phys. A \textbf{32} (1999), 2341-2359
%doi:10.1088/0305-4470/32/12/008
[arXiv:cond-mat/9807076 [cond-mat.stat-mech]].

\bibitem{Dunning:2002tt}
C.~Dunning,
%``Finite size effects and the supersymmetric sine-Gordon models,''
J. Phys. A \textbf{36} (2003), 5463-5476
%doi:10.1088/0305-4470/36/20/308
[arXiv:hep-th/0210225 [hep-th]].

\bibitem{Kulish:2004ap}
P.~P.~Kulish and A.~M.~Zeitlin,
%``Superconformal field theory and SUSY N=1 KdV hierarchy. 1. Vertex operators and Yang-Baxter equation,''
Phys. Lett. B \textbf{597} (2004), 229-236
%doi:10.1016/j.physletb.2004.07.019
[arXiv:hep-th/0407154 [hep-th]].

\bibitem{Kulish:2005qc}
P.~P.~Kulish and A.~M.~Zeitlin,
%``Superconformal field theory and SUSY N=1 KDV hierarchy II: The Q-operator,''
Nucl. Phys. B \textbf{709} (2005), 578-591
%doi:10.1016/j.nuclphysb.2004.12.031
[arXiv:hep-th/0501019 [hep-th]].

\bibitem{Bazhanov:1994ft}
V.~V.~Bazhanov, S.~L.~Lukyanov and A.~B.~Zamolodchikov,
%``Integrable structure of conformal field theory, quantum KdV theory and thermodynamic Bethe ansatz,''
Commun. Math. Phys. \textbf{177} (1996), 381-398
%doi:10.1007/BF02101898
[arXiv:hep-th/9412229 [hep-th]].

\bibitem{Bena:2003wd}
I.~Bena, J.~Polchinski and R.~Roiban,
%``Hidden symmetries of the AdS(5) x S**5 superstring,''
Phys. Rev. D \textbf{69} (2004), 046002
%doi:10.1103/PhysRevD.69.046002
[arXiv:hep-th/0305116 [hep-th]].

\bibitem{Sun:2012xw}
J.~Sun,
%``Polynomial relations for $q$-characters via the ODE/IM correspondence,''
SIGMA \textbf{8} (2012), 028
%doi:10.3842/SIGMA.2012.028
[arXiv:1201.1614 [math.QA]].

\bibitem{Masoero:2015lga}
D.~Masoero, A.~Raimondo and D.~Valeri,
%``Bethe Ansatz and the Spectral Theory of Affine Lie Algebra-Valued Connections I. The simply-laced Case,''
Commun. Math. Phys. \textbf{344} (2016) no.3, 719-750
%doi:10.1007/s00220-016-2643-6
[arXiv:1501.07421 [math-ph]].

\bibitem{Ito:2020htm}
K.~Ito, T.~Kondo, K.~Kuroda and H.~Shu,
%``ODE/IM correspondence for affine Lie algebras: A numerical approach,''
J. Phys. A \textbf{54} (2021) no.4, 044001
%doi:10.1088/1751-8121/abd21e
[arXiv:2004.09856 [hep-th]].

\bibitem{serganova}
	V.~V.~Serganova,
	Math. USSR Izvestiya, {\bf 24} (1985) 539.

\bibitem{Leites:1985hh}
D.~A.~Leites, M.~V.~Saveliev and V.~V.~Serganova,
%``EMBEDDINGS OF LIE SUPERALGEBRA OSP(1/2) AND THE ASSOCIATED NONLINEAR SUPERSYMMETRIC EQUATIONS,''
%IFVE-85-81.
 in Group-theoretical methods in Physics, Proceedings of the Third Yurmala Seminar (Yurmala, 1995), Moscow, 1986, 255-298.
\bibitem{Kac} V.G.~Kac, Adv. Math.{\bf 26} (1977) 8.
%A Sketch of Lie Superalgebra Theory, Comm. Math. Phys. 53 (1977) 31-64

\bibitem{FSS} L. Frappat, A. Sciarrino and P. Sorba,
%1989 Structure of Basic Lie Superalgebras and of their Affine Extensions. 
Commun. Math. Phys. 121 (1989), 457-500 %doi:10.1007/BF01217734

\bibitem{vanderleur}
J.W. van der Leur, Commun. Algebra {\bf 18} (1989) 1815.

\bibitem{Olshanetsky:1982sb}
M.~A.~Olshanetsky,
%``SUPERSYMMETRIC TWO-DIMENSIONAL TODA LATTICE,''
Commun. Math. Phys. \textbf{88} (1983), 63
%doi:10.1007/BF01206879

\bibitem{Evans:1990qq}
J.~Evans and T.~J.~Hollowood,
%``Supersymmetric Toda field theories,''
Nucl. Phys. B \textbf{352} (1991), 723-768
[erratum: Nucl. Phys. B \textbf{382} (1992), 662-662]
%doi:10.1016/0550-3213(91)90105-7

\bibitem{Komata:1990cb}
S.~Komata, K.~Mohri and H.~Nohara,
%``Classical and quantum extended superconformal algebra,''
Nucl. Phys. B \textbf{359} (1991), 168-200
%doi:10.1016/0550-3213(91)90296-A

\bibitem{Delduc:1991sg}
F.~Delduc, E.~Ragoucy and P.~Sorba,
%``SuperToda theories and W algebras from superspace Wess-Zumino-Witten models,''
Commun. Math. Phys. \textbf{146} (1992), 403-426
%doi:10.1007/BF02102635

\bibitem{Andreev:1987xj}
V.~A.~Andreev,
%``Odd Bases of Lie Superalgebras and Integrable Equations,''
Theor. Math. Phys. \textbf{72} (1987), 758-764
%doi:10.1007/BF01035702

\bibitem{Scheunert:1976wj}
M.~Scheunert, W.~Nahm and V.~Rittenberg,
%``Irreducible Representations of the OSP(2,1) and SPL(2,1) Graded Lie Algebras,''
J. Math. Phys. \textbf{18} (1977), 155
%doi:10.1063/1.523149

\bibitem{Marcu:1979se}
M.~Marcu,
%``The Representations of Spl(2,1): An Example of Representations of Basic Superalgebras,''
J. Math. Phys. \textbf{21} (1980), 1277
%doi:10.1063/1.524576

\bibitem{Gotz:2005jz}
G.~Gotz, T.~Quella and V.~Schomerus,
%``Representation theory of sl(2|1),''
J. Algebra \textbf{312} (2007), 829-848
%doi:10.1016/j.jalgebra.2007.03.012
[arXiv:hep-th/0504234 [hep-th]].

\bibitem{Dorey:2007zx}
P.~Dorey, C.~Dunning and R.~Tateo,
%``The ODE/IM Correspondence,''
J. Phys. A \textbf{40} (2007), R205
%doi:10.1088/1751-8113/40/32/R01
[arXiv:hep-th/0703066 [hep-th]].

\bibitem{Tsuboi:1997iq}
Z.~Tsuboi,
%``Analytic Bethe ansatz and functional equations for Lie superalgebra $sl(r+1|s+1)$,''
J. Phys. A \textbf{30} (1997), 7975-7991
%doi:10.1088/0305-4470/30/22/031
[arXiv:0911.5386 [math-ph]].


\bibitem{Tsuboi:1999ft}
Z.~Tsuboi,
%``Analytic Bethe Ansatz and functional relations related to tensor like representations of type-II Lie superalgebras B(r|s) and D(r|s),''
J. Phys. A \textbf{32} (1999), 7175-7206
%doi:10.1088/0305-4470/32/41/311
[arXiv:0911.5393 [math-ph]].

\end{thebibliography}
\end{document}